\documentclass{article}
\usepackage[preprint]{neurips_2023}
\usepackage{amsmath,amssymb,amsfonts}
\usepackage{algorithm,algorithmic}
\usepackage{graphicx}
\usepackage{textcomp}
\usepackage{xcolor}
\def\BibTeX{{\rm B\kern-.05em{\sc i\kern-.025em b}\kern-.08em
    T\kern-.1667em\lower.7ex\hbox{E}\kern-.125emX}}

\usepackage{booktabs}
\usepackage{multirow}
\usepackage{tcolorbox}
\newtcolorbox{ansbox}{colframe = gray!75!black}
\usepackage{makecell}
\usepackage{colortbl}
\usepackage{subcaption}
\usepackage{ulem}
\usepackage{paralist}

\usepackage{url}
\usepackage{hyperref}

\newcommand{\colorit}{\cellcolor{gray!25}}

\newcommand{\mygray}{gray!25}

\makeatletter
\def\@onedot{\ifx\@let@token.\else.\null\fi\xspace}

\makeatother

\begin{document}

\newcommand{\abb}{MORTAR}
\newcommand{\todo}[1]{\textbf{\textcolor{red}{TODO: #1}}}
\newcommand{\js}[1]{\noindent\textcolor{blue}{(JS): #1}}
\newcommand{\zz}[1]{\textbf{\textcolor{green}{Zhehua: #1}}}

\title{\abb: A Model-based Runtime Action Repair Framework for AI-enabled Cyber-Physical Systems
}

\author{%
  Renzhi Wang \\
  University of Alberta\\
  \texttt{renzhi.wang@ualberta.ca} \\
  \And
  Zhehua Zhou\\
  University of Alberta\\
  \texttt{zhehua1@ualberta.ca} \\
  \And
  Jiayang Song\\
  University of Alberta\\
  \texttt{jiayan13@ualberta.ca} \\
  \And
  Xuan Xie\\
  University of Alberta\\
  \texttt{xxie9@ualberta.ca} \\
  \And
  Xiaofei Xie\\
  Singapore Management University\\
  \texttt{xfxie@smu.edu.sg}\\
  \And
  Lei Ma\\
  The University of Tokyo \\ University of Alberta\\
  \texttt{ma.lei@acm.org} \\
}

\maketitle

\begin{abstract}
Cyber-Physical Systems (CPSs) are increasingly prevalent across various industrial and daily-life domains, with applications ranging from robotic operations to autonomous driving.
With recent advancements in artificial intelligence (AI), learning-based components, especially AI controllers, have become essential in enhancing the functionality and efficiency of CPSs. 
However, the lack of interpretability in these AI controllers presents challenges to the safety and quality assurance of AI-enabled CPSs (AI-CPSs). 
Existing methods for improving the safety of AI controllers often involve neural network repair, which requires retraining with additional adversarial examples or access to detailed internal information of the neural network. 
Hence, these approaches have limited applicability for black-box policies, where only the inputs and outputs are accessible during operation.
To overcome this, we propose \abb{}, a runtime action repair framework designed for AI-CPSs in this work. 
\abb{} begins by constructing a prediction model that forecasts the quality of actions proposed by the AI controller. 
If an unsafe action is detected, \abb{} then initiates a repair process to correct it. 
The generation of repaired actions is achieved through an optimization process guided by the safety estimates from the prediction model. 
We evaluate the effectiveness of \abb{} across various CPS tasks and AI controllers.
The results demonstrate that \abb{} can efficiently improve task completion rates of AI controllers under specified safety specifications. 
Meanwhile, it also maintains minimal computational overhead, ensuring real-time operation of the AI-CPSs.

\end{abstract}

\section{Introduction}\label{sec:intro}

Cyber-Physical Systems (CPSs) are hybrid systems that integrate computing units and mechanical components. The digital and physical elements actively interact to perform target tasks collaboratively in complex environments. 
In recent years, research and development of CPSs have grown rapidly across various domains, including robotics~\cite{nikolakis2019cyber}, autonomous driving~\cite{jia2021integrated} and power systems~\cite{yohanandhan2020cyber}.
With the nature of interactive cyber and physical interactions, CPSs are considered to initiate leading-edge performance compared to traditional embedded systems with improved efficiency, reliability and robustness~\cite{castiglioni2024stark, tabuada2014towards, rungger2015notion}.

With the development in the field of Artificial Intelligence (AI), learning-based approaches retain problem-solving abilities across various applications. Thus, many CPSs have adopted AI-powered components in their solutions to enhance their adaptability and generalizability.

These AI-powered systems are usually referred to as AI-enabled CPSs (AI-CPSs). 
Among the applications development, a key focus is how to use Deep Reinforcement Learning (DRL) to create control policies to better handle the complex system dynamics of modern CPSs~\cite{plaat2020deep,liu2019reinforcement}.

In contrast to the conventional control policy design, which typically relies on control-theoretical concepts, such as Proportional Integral Derivative (PID)~\cite{ang2005pid} and Model Predictive Control (MPC)~\cite{morari1988model}, DRL-based controllers learn control policies through interactions with the environment and thus do not require the explicit solving of complex system dynamics. 
Therefore, such AI-CPSs\footnote{In this work, we refer to AI-CPSs as CPSs with DRL control policies.} generally exhibit enhanced performance and adaptability in complex systems and possess high generalizability across various application scenarios~\cite{radanliev2021artificial,song2022cyber,lv2021artificial,xie2023mosaic,song2023mathtt,hu2023AIoTML}.

\begin{figure*}[t]
    \centering
    \includegraphics[width=0.95\linewidth]{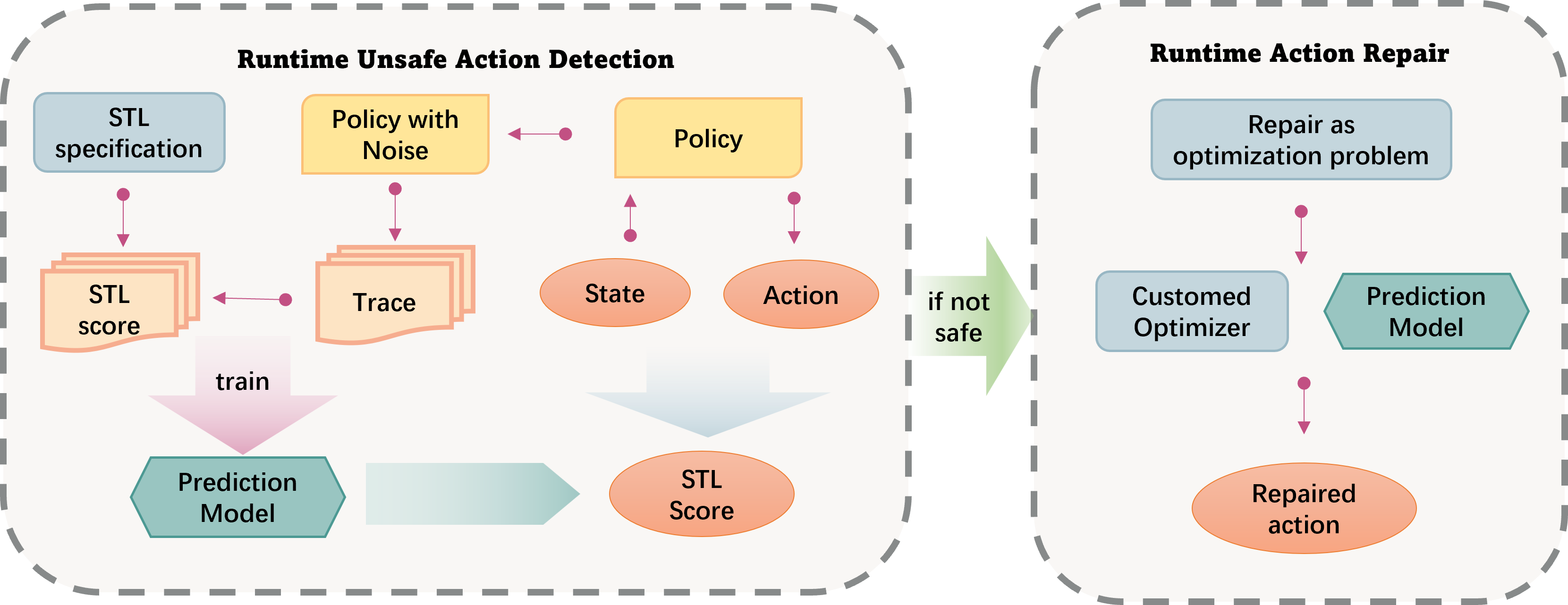}
    \caption{Overview of \abb{}}
    \label{fig:workflow}
\end{figure*}

Despite their extensive capabilities, learning-based control policies still meet several distinctive challenges.
For instance, the lack of interpretability and formal verification could lead these control policies to produce faulty actions in specific scenarios. The failure may be caused by biased training data, insufficient training, or noisy operations.
Furthermore, unlike conventional control systems, providing safety guarantees to learning-based control policies using expert knowledge or mathematical analysis is often challenging. 
As a result, it is extremely complex to ensure the safety of AI-CPSs~\cite{li2020early}.

To address this issue, previous works propose various solutions, and a common technical route among these approaches to enhance the safety of learning-based control policies is through neural network (NN) repair~\cite{yang2022neural, lyu2023autorepair}, which aims to enhance NN performance via adversarial training or parameter alternation.
Existing repair strategies mainly focus on safeguarding the decision-making logic during training or augmenting the training data by removing unreliable samples that could lead to faulty control policies~\cite{lyu2023autorepair}.
However, these approaches typically apply only to white-box policies where the internal states of the neural networks are known and modifiable.
Consequently, these approaches are hard to apply to black-box policies that involve networks with unknown details or are immutable.
Moreover, these retraining methods often also incur huge computational costs.
Given that CPSs are widely deployed in safety-critical settings, any unsafe control command can lead to dangerous behaviour, which can cause failure and even harm.
These challenges greatly hinder the further development and deployment of AI-CPSs across various domains.

Ideally, an effective repair scheme for AI controllers should be compatible with black-box policies, facilitate real-time action correction, and be easily transferable to various CPSs at low cost.
Inspired by this, we propose in this work \textbf{\abb{}}, a model-based runtime action repair framework for AI-CPSs.
Instead of requiring detailed neural network information, \abb{} utilizes only the input and output data from the AI controller to rectify unsafe actions in real-time. 
It consists of two parts: \textit{prediction model construction} and \textit{repaired action generation}. 
Given an AI-CPS and a DRL control policy that needs to be repaired, we first collect data on the safety of various state-action pairs by executing the specified DRL policy. 
Then, leveraging this collected data, we construct a prediction model that estimates the quality of actions generated by the DRL policy based on safety-related system specifications. 
Consequently, if an unsafe action is identified by the prediction model, \abb{} employs a gradient-based optimization technique to immediately generate a repaired action that satisfies the safety requirements.
See Fig.~\ref{fig:workflow} for an overview of \abb{}.

To evaluate the effectiveness of \abb{}, we design experiments to compare our method with a commonly used action-level repair strategy in the domain of safe reinforcement learning.
Specifically, a large-scale evaluation is performed with five representative CPSs, eight system safety specifications, ten DRL controllers, and over 300 experiment runs. 
The results demonstrate that our constructed prediction model can accurately track the quality of actions with respect to system safety specifications, and the proposed \abb{} framework outperforms the baseline method by generating repaired safe actions more effectively and reliably. 
Moreover, compared to existing methods, \abb{} also achieves a better balance between the repair effectiveness and the computational cost.
The contributions of this work are summarized as follows:

\begin{compactitem}[$\bullet$]
    \item We propose a novel method to construct a prediction model that evaluates the quality of controller outputs at runtime. 
    The prediction model behaves as a runtime monitor to track the safety of the system and initiate a repair procedure when needed.
    \item We introduce a black-box action-level repair strategy that leverages the prediction model to generate safer actions in real-time using a gradient-based optimization technique.
    \item A large-scale experiment is performed to investigate the effectiveness of our proposed method. 
    The results confirm that \abb{} surpasses existing repair methods and is able to produce safer and more reliable control outputs. 
\end{compactitem}


\noindent \textbf{The Contribution to the Software Engineering Field.}
CPSs are one of the core research directions in the field of Software Engineering (SE), as they are promising candidates to bridge the cyber and physical worlds.
The operational behavior of CPSs is primarily determined by their software components, making quality assurance for CPSs even more critical.
Our work is dedicated to safeguarding the quality of AI-CPSs by establishing a general framework to amend the outputs of AI controllers at runtime.
Towards quality-assured AI-CPSs, we hope our work can enhance and spread software-defined CPSs across more practical applications in diverse domains.

The rest of the paper is structured as follows.
Section~\ref{sec:background} introduces the corresponding backgrounds. Section~\ref{sec:approach} details the construction of the prediction model and the runtime action-level repair strategy of \abb{}. 
Section~\ref{sec:evaluation} designs a series of experiments to explore the effectiveness of \abb{}. 
Section~\ref{sec:discuss} discusses several key findings from our experiments. 
Section~\ref{sec:threats} analyzes potential threats and limitations associated with \abb{}, while Section~\ref{sec:relatedWork} reviews related works. 
Finally, Section~\ref{sec:conclusion} concludes the paper.

\section{Background}\label{sec:background}
In this section, we provide essential background knowledge about AI-CPSs and Signal Temporal Logic (STL), which is commonly used to define system safety specifications.

\subsection{AI-enabled Cyber-Physical Systems}

\begin{figure}[t]
    \centering
    \includegraphics[width=0.65\linewidth]{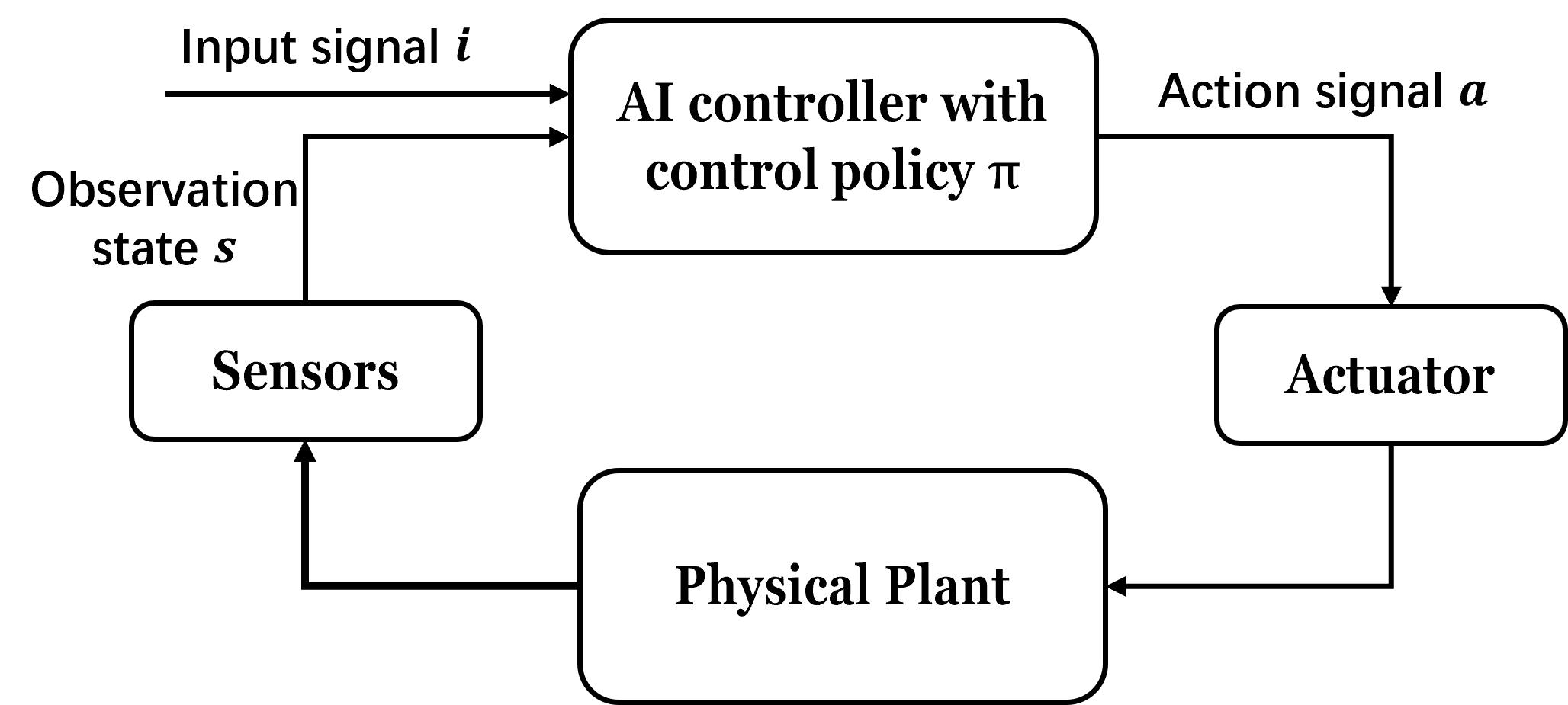}
    \caption{Workflow for a CPS with an AI controller    }
    \label{fig:cps}
    \vspace{-10pt}
\end{figure}

CPSs integrate computational elements with physical processes ~\cite{lee2015past,alguliyev2018cyber}. 
A typical CPS consists of four major components (see Fig.~\ref{fig:cps}): a physical plant $E$, a network of sensors, an actuator and a control unit~\cite{song2016cyber,song2022cyber,afzal2020study}.
At a given timestep $t$, the controller receives the current system state $s_t$ through sensors and, in response to an input signal $i$, formulates an action $a_t$. 
The actuator then executes this action, causing the physical plant $E$ to transition to a subsequent state in accordance with the system dynamics~\cite{song2016cyber}. 

For AI-CPSs considered in this work, the control unit is implemented as a DRL-based control policy $\pi$. 
Recent advancements in DRL have introduced several high-performing algorithms, such as Deep Deterministic Policy Gradient (DDPG)~\cite{lillicrap2015continuous}, Proximal Policy Optimization (PPO)~\cite{schulman2017proximal}, and Trust Region Policy Optimization (TRPO)~\cite{schulman2015trust}, which show considerable promise as AI controllers for CPSs.
Compared to traditional controllers that rely on control-theoretical concepts, e.g., PID or MPC, DRL-based controllers circumvent explicit solutions for system dynamics, thereby enabling a better performance and generalizability in systems with complex dynamics~\cite{lin2020comparison,liu2019reinforcement,leong2020deep,Hu2020Quantitative}. 
However, limited by the interpretability and explainability, the actions generated by DRL controllers might not adhere to system safety specifications, leading to potentially hazardous behaviors. 
Hence, in this work, we aim to enhance the safety performance of DRL controllers by using \abb{} to rectify their generated actions to align with given safety requirements.

\subsection{Signal Temporal Logic (STL)}\label{sec:background_stl}

STL~\cite{donze2010robust} is a specification language designed to describe the expected temporal behavior of a system. 
It extends the Linear Temporal Logic (LTL) by supporting continuous-time frameworks and real-valued signals. 
This makes it particularly valuable in domains where continuous time analysis of system behaviors is essential, such as environmental monitoring and system control~\cite{bartocci2018specification}.

STL is widely used to define system safety specifications.
It adopts temporal operators similar to those found in LTL, such as \textit{Globally} (G), \textit{Finally} (F) and \textit{Eventually} (E). 
For example, a typical STL specification using the \textit{Globally} operator is $G_{[a,b]}(x<\theta)$, which indicates that throughout the interval from time $a$ to time $b$, the signal $x$ must remain to blow a threshold $\theta$.
This capability to specify conditions over continuous time intervals allows STL  to offer detailed and quantitative safety specifications regarding the temporal behavior of a system.

In this work, we employ STL to define safety requirements.
Meanwhile, to quantitatively measure the degree of satisfaction of a given safety specification, we utilize the \textit{quantitative robust semantics}~\cite{donze2010breach}.
Given an STL specification $\varphi$ and a system output trajectory $\zeta$, the quantitative robust semantics $\textsc{rob}(\zeta, \varphi)$ maps $\zeta$ and $\varphi$ to a real number. 
A positive or negative value of $\textsc{rob}(\zeta, \varphi)$ indicates whether the specification is satisfied or violated, and a larger value implies a stronger satisfaction or violation.
We refer to this value as the \textit{STL score} of the trajectory $\zeta$ under the specification $\varphi$ in this paper and utilize it to construct the prediction model. 
Further details are provided in Section~\ref{sec:approach_monitor}.

\section{Approach}\label{sec:approach}

\begin{figure*}[ht]
    \centering
    \includegraphics[width=0.9\linewidth]{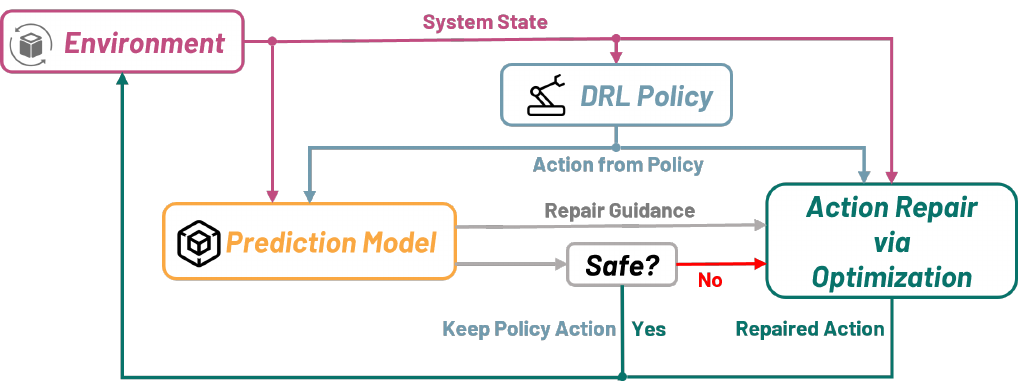}
    \caption{Runtime workflow of \abb{}}
    \vspace{-5pt}
    \label{fig:method_overview}
\end{figure*}

In this section, we first introduce the problem formulation considered in this work (Section~\ref{sec:approach_problem_fromulation}).
Then, we present details about our proposed framework \abb{}, which conducts action-level repair for AI-CPSs.
Through collecting data on the safety of various state-action pairs, we construct a prediction model that estimates the safety of outputs of the DRL control policy (Section~\ref{sec:approach_monitor}).
Using this constructed prediction model, we identify actions that may lead to unsafe behaviors. 
Subsequently, an optimization-based technique is employed to generate repaired actions during runtime, enhancing the safety performance of the DRL control policy (Section~\ref{sec:approach_repair}). 
See also Fig.~\ref{fig:method_overview} for an illustration of the runtime workflow of \abb{}.
Further details are provided as follows. 

\subsection{Problem Formulation}\label{sec:approach_problem_fromulation}

In this work, we consider the scenario where a CPS and a DRL control policy $\pi$ is given. 
At each timestep $t$, the policy $\pi$ receives the system state $s_t$ and outputs an action $a_t = \pi(s_t)$.
The CPS then evolves to the subsequent state $s_{t+1}$ by executing the action $a_t$.
Running the system under policy $\pi$ for a horizon of $T$ timesteps results in a system trajectory $\zeta = (s_0,s_{1},\ldots,s_{T})$, where $s_0$ is the initial system state.
For a given safety specification $\varphi$, we aim to ensure that the system trajectory $\zeta$ satisfies the specification $\varphi$, i.e., the corresponding STL score $\textsc{rob}(\zeta, \varphi)$ should be positive.

However, due to various factors such as insufficient training or environmental noise, the DRL control policy $\pi$ may generate faulty actions that fail to meet the safety specification $\varphi$.
Therefore, the objective of \abb{} is to correct potential erroneous actions in runtime by using only the input and output of the control policy $\pi$, i.e., the state $s_t$ and action $a_t$.
This correction is facilitated through an action-level repair strategy, where at each timestep $t$, we first predict whether the proposed action $a_t$ will lead to a future trajectory $\zeta$ that violates the safety specification $\varphi$.
If an action $a_t$ is deemed unsafe, we then generate a repaired action $a_{t}^r$ that enhances the safety. 
Details about \abb{} are explained in the following subsections.

\subsection{Construction of the Prediction Model}\label{sec:approach_monitor}

The first and critical step of \abb{} is to construct a prediction model that identifies potential risky behaviours of the given DRL control policy. 
As mentioned in Section~\ref{sec:background_stl}, we utilize STL to define the safety requirements for AI-CPS and employ the STL score to assess how well the given system output trajectory satisfies these safety constraints.
However, the STL score cannot be directly calculated for future system behaviors where the precise system output trajectory is unknown.
To address this issue, we therefore introduce a prediction model that estimates the future STL score for the current system state $s_t$ and action $a_t$ under the existing DRL control policy $\pi$.
Constructing this prediction model involves two steps: data collection and model training, which are detailed as follows.

\subsubsection{Data Collection}\label{sec:approach:data_collection}

We employ the specified control policy $\pi$ to gather training data for constructing the prediction model. 
Note that if the control policy $\pi$ is already well-trained, the collected data will predominantly consist of safe trajectories, with unsafe trajectories constituting only a small fraction.
This imbalance can hinder the development of an efficient prediction model. 
To address this issue, we introduce random action noises into the control policy $\pi$ during the data collection phase to balance the proportion of safe and unsafe trajectories (see Section~\ref{sec:experiment_setting_rq1} and Table~\ref{table:collected_stats} for an example).
This strategy facilitates the construction of a more accurate prediction model capable of identifying potential unsafe actions.

By executing the control policy $\pi$ with added noises, we collect a set of safe and unsafe trajectories, where each trajectory $\zeta$ contains $T$ timesteps, and each timestep corresponds to a state-action pair $\langle s_t, a_t \rangle$. 
The STL score $\textsc{rob}(\zeta, \varphi)$ for each trajectory can then be calculated using the specification $\varphi$.
Considering the strong temporal correlation within each trajectory, we assign the same computed STL score $\textsc{rob}(\zeta, \varphi)$ to every state-action pair within that trajectory. 
This obtained dataset of state-action pairs, labeled with their corresponding STL scores, is then used to train the prediction model.

\subsubsection{Prediction Model Training}\label{sec:prediction_model}

To predict the safety of each proposed action $a_t= \pi(s_t)$ during runtime, we train a prediction model $M$ using the collected dataset. 
For a given state-action pair $\langle s_t, a_t \rangle$, the prediction model $M$ estimates the future STL score $\textsc{rob}(\zeta, \varphi)$ for that pair. 
For brevity, we denote the output of the prediction model $M$, i.e., the estimated STL score, as $\psi_t = M(s_t,a_t)$ throughout the remainder of this paper. 
In this work, the prediction model $M$ is represented by a lightweight neural network and is specifically trained for each CPS task under consideration.

In \abb{}, the prediction model $M$ serves as an online monitor that assesses the safety of actions generated by the control policy $\pi$ in real-time. 
By introducing a predefined threshold $\psi_{thres}\geq 0$, we classify each proposed action $a_t$ as safe if the output of the prediction model satisfies that $\psi_t = M(s_t,a_t) \geq \psi_{thres}$. 
If the output falls below the threshold, i.e., $\psi_t = M(s_t,a_t) < \psi_{thres}$, the action $a_t$ is considered unsafe, triggering a subsequent action repair process. 
Note that these unsafe actions may not indicate immediate failures but could lead to potential constraint violations in future steps. 
Moreover, rather than a binary classifier, the prediction model $M$ provides a continuous value in terms of safety, which greatly facilitates the action repair process by enabling the estimation of gradient information to guide improvements in action safety.

\subsection{Repaired Action Generation}\label{sec:approach_repair}

With the constructed prediction model, we are able to predict the safety of generated actions at each timestep. 
Once an unsafe action is identified, we initiate the action repair process, which forms the second component of \abb{}. 
In this subsection, we provide details on how to compute a repaired action using the safety estimates obtained from the prediction model.

\subsubsection{Action Repair as An Optimization Problem}

After determining an unsafe action $a_t$, we aim to rectify it by identifying a repaired action $a_t^r$ that satisfies $M(s_t,a_t^r)  \geq \psi_{thres}$. 
We define the repaired action $a_t^r$ as follows
\begin{align}
    a_t^r = a_t+a_t^p
\end{align}
where $a_{t}^p$ is the repair patch that needs to be calculated. 

By using the prediction model $M$, the repaired action $a_t^r$ can be computed by solving the following optimization problem
\begin{align}
    \min_{a_t^p} &||M(s_t,a_t+a_t^p) - \psi_{max}||^2 \label{eq:minimize_goal}\\
    \text{s.t.} \quad & M(s_t,a_t+a_t^p)  \geq \psi_{thres} \label{eq:minimize_thres}\\
        &M(s_t,a_t+a_t^p) \leq \psi_{max}\label{eq:minimize_psi} \\
        &A_{min} \leq a_t+a_t^p \leq A_{max} \label{eq:minimize_action}        
\end{align}
The objective~\eqref{eq:minimize_goal} of this optimization is to minimize the difference between the safety estimate from the prediction model $M$ and a maximum value $\psi_{max}$, where we have $\psi_{max} > \psi_{thres}$.
The rationale behind this is that, due to the well-acknowledged balance between ensuring safety and encouraging exploration for higher rewards~\cite{garcia2015comprehensive,gu2022review}, optimizing the output $M(s_t,a_t+a_t^p)$ towards an excessively high value may result in actions that are overly restrictive for safety, potentially neglecting the completion of task requirements. 
Therefore, instead of directly maximizing the output $M(s_t,a_t+a_t^p)$, i.e., finding the safest action, we aim to keep it close to the maximum value $\psi_{max}$ during the optimization process.
In this work, we calculate the maximum value $\psi_{max}$ as follows
\begin{align}
    \psi_{max} = \Bar{\psi} + 2\sigma_\psi
    \label{eq:maximum_psi}
\end{align}
where $\Bar{\psi}$ and $\sigma_\psi$ are the mean value and standard deviation of the STL scores $\textsc{rob}(\zeta, \varphi)$ for all trajectories collected in the training dataset for the prediction model, respectively.

While constraint~\eqref{eq:minimize_thres} ensures that the identified repaired action $a_t^r = a_t+a_t^p$ is deemed safe by the prediction model, constraint~\eqref{eq:minimize_psi} indicates that the output should not exceed the maximum value $\psi_{max}$, thus preventing unreasonable high values in safety estimates.
Constraint~\eqref{eq:minimize_action} ensures that the computed repaired action $a_t^r$ remains within the valid action space and avoids outliers.
The boundaries $A_{min}$ and $A_{max}$ are calculated as follows
\begin{align}
    A_{min} = \max(\Bar{A}-2\sigma_A, a_{min})\\
    A_{max} = \min(\Bar{A}+2\sigma_A, a_{max})
\end{align}
where $\Bar{A}$ and $\sigma_A$ are, respectively, the mean value and standard deviation of all actions observed in the training dataset.
$a_{min}$ and $a_{max}$ are the minimum and maximum actions recorded across the entire dataset.

By solving the optimization problem~\eqref{eq:minimize_goal}-\eqref{eq:minimize_action}, we determine a repaired action $a_t^r = a_t + a_t^p$ that offers enhanced safety. 
The proposed \abb{} framework, as detailed in Algorithm~\ref{algo:monitor_repair} and Fig.~\ref{fig:method_overview}, therefore utilizes this repair process to facilitate safer operation of the AI-CPS.

\begin{algorithm}[t]
 \caption{\abb{}}
 \label{algo:monitor_repair}
\begin{algorithmic}[1]
        \REQUIRE CPS plant $E$, prediction model $M$, DRL control policy $\pi$, threshold $\psi_{thres}$
        \WHILE{$E.active$}
            \STATE $s_t \gets E.observsation$
            \STATE $a_t \gets \pi(s_t) $
            \STATE $\psi_t \gets M(s_t,a_t)$
            \IF{$\psi_t \geq \psi_{thres}$}
                \STATE $a_t^r = a_t$
            \ELSE
                \STATE $a_t^r \gets \text{Optimization Solving~\eqref{eq:minimize_goal}-\eqref{eq:minimize_action}}$
            \ENDIF
            \STATE $E.step(s_t, a_t^r)$
        \ENDWHILE
    \end{algorithmic}
\end{algorithm}

\subsubsection{Practical Optimization Solving}
The major challenge in solving the optimization~\eqref{eq:minimize_goal}-\eqref{eq:minimize_action} is ensuring a time-efficient computation to minimize overhead in the control operation. 
Considering this, traditional search-based methods, such as Nelder-Mead~\cite{gao2012implementing}, and gradient-based methods, such as BFGS~\cite{bazaraa2013nonlinear}, often fail to provide a satisfactory performance. 
These methods may require a considerable amount of iterations to find a feasible solution or need to compute estimates of the Hessian matrix for gradients, both of which are computationally expensive.

The key to reducing computational overhead lies in minimizing the number of times the prediction model is used to calculate predictions. 
Motivated by this, we utilize the targeted Basic Iterative Method (BIM)~\cite{xiao2018generating} in this work to determine the repaired action $a_t^r$.
BIM is a widely adopted method in generating adversarial examples~\cite{goodfellow2014explaining,kurakin2018adversarial,yuan2019adversarial}.
It aims to incrementally modify the input so that the neural network produces a specific, altered classification result. 
Similarly, the action repair process seeks to modify actions to achieve a different output from the prediction model, paralleling the objectives of adversarial example generation and thereby motivating the use of the BIM method.

By setting the target to the maximum value $\psi_{max}$ from~\eqref{eq:maximum_psi}, the targeted BIM attempts to find an adversarial action patch $a_{adv} = \text{BIM}(M, s_t, a_t, \psi_{max}, \psi_{thres})$ that moves the output of the prediction model towards $\psi_{max}$. 
Note that rather than approaching $\psi_{max}$ as closely as possible, the primary objective of the action repair process is to identify an action that is considered safe by the prediction model.
Therefore, to further enhance computational efficiency, we incorporate an early-stop strategy within the BIM search process. 
Specifically, the search is terminated when the repaired action $a_t^r = a_t + a_{adv}$ yields a value from the prediction model that falls within the range between $\psi_{thres}$ and $\psi_{max}$.

It is also important to note that, unlike traditional adversarial example generation, where classification results can change immediately, controlling CPSs generally requires smoother command inputs. 
Large and sudden changes in actions may lead to system instability or excessive oscillating behavior~\cite{zhou2020runtime}. 
To address this, we additionally introduce a step size $\alpha$ to control the magnitude of the identified adversarial action patch $a_{adv}$, i.e., we have $a_t^p = \alpha * a_{adv}$. 
Moreover, the computed repaired action $a_t^r$ must be constrained by $A_{min}$ and $A_{max}$ to ensure it remains within valid action boundaries. 
Accordingly, the complete process for solving the optimization~\eqref{eq:minimize_goal}-\eqref{eq:minimize_action} using targeted BIM is detailed in Algorithm~\ref{algo:optimize solving}. 
The proposed algorithm is able to find a solution in a limited number of iterations while maintaining satisfactory quality, thereby achieving a better balance between performance and computational cost.

\begin{algorithm}[t]
 \caption{Optimzation Solving}
 \label{algo:optimize solving}
\begin{algorithmic}[1]
        \REQUIRE Prediction model $M$, State $s_t$, Action $a_t$, Step size $\alpha$, Action space $[A_{min},A_{max}]$, Maximum value $\psi_{max}$, Threshold $\psi_{thres}$
        \STATE $a_{adv} = \text{BIM}(M, s_t, a_t, \psi_{max}, \psi_{thres})$
        \STATE $a_t^p = \alpha * a_{adv}$
        \STATE $a_t^r = a_t + a_t^p$ 
        \STATE $a_t^r \gets [A_{min},A_{max}] $
    \end{algorithmic}
\end{algorithm}

\section{Experimental Evaluation}\label{sec:evaluation}

To evaluate the effectiveness of \abb{}, we conduct comprehensive experiments across various tasks. 
This section presents details about our research questions, the systems studied, the experimental setup, and the corresponding evaluation results.

\begin{figure*}[t]
  \centering
  \vspace{5pt}
  \subfloat[Point Reaching]{\includegraphics[height=3cm]{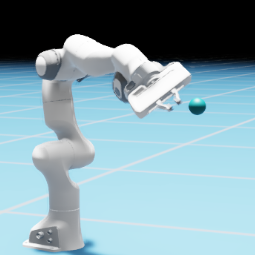}}
  \hspace{2mm}
  \subfloat[Cube Stacking]{\includegraphics[height=3cm]{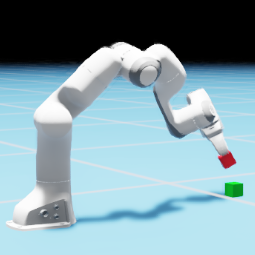}}
  \hspace{2mm}
  \subfloat[Peg In Hole]{\includegraphics[height=3cm]{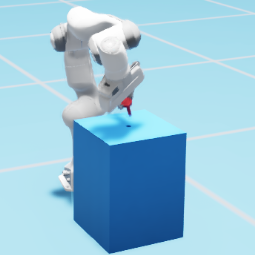}}
  \hspace{2mm}\\
  \subfloat[Ball Balancing]{\includegraphics[height=3cm]{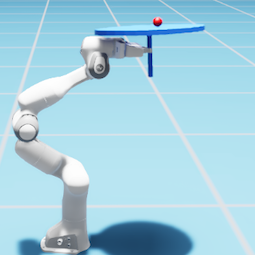}}
  \hspace{2mm}
  \subfloat[Ball Catching]{\includegraphics[height=3cm]{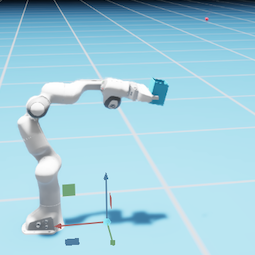}}
   
  \caption{Five robotic manipulation tasks simulated with NVIDIA Omniverse Isaac Sim.}
  \label{fig:task_demo}
\end{figure*}

\subsection{Research Questions}
We aim to investigate the following research questions (RQs) for assessing the effectiveness of \abb{} from various perspectives.

\newcommand{\rqone}{RQ1: To what extent can the prediction model accurately predict the system safety?}
\newcommand{\rqtwo}{RQ2: How well does \abb{} perform in the studied CPSs?}
\newcommand{\rqthree}{RQ3: How do different optimization algorithms impact the performance of \abb{}?}

\begin{compactitem}[$\bullet$]

\item\noindent\textbf{\rqone}

The prediction model plays a vital role in our online repair scheme; namely, it determines whether to initiate the repair process and provides guidance for the generation of repaired actions.
Therefore, this RQ aims to investigate whether the constructed prediction model can precisely estimate the future STL sores of the subject CPSs.

\item\noindent\textbf{\rqtwo{}} 

In this RQ, we compare the performance of \abb{} with the original baseline policy and another representative action repair method, which is referred to as Safe Exploration (SafeExp)~\cite{dalal2018safe} in this paper.
The performance is evaluated from two perspectives: (1) the effectiveness of the repair method in enhancing the safety of the original policy and (2) the impact of the repair method on computational overhead, ensuring it does not significantly interfere with system operations.

\item\noindent\textbf{\rqthree{}}\label{sec:propose_rq3}

A critical factor that can notably affect the performance of \abb{} is the choice of the algorithm used to solve the optimization problem~\eqref{eq:minimize_goal}-\eqref{eq:minimize_action}.
Hence, in this RQ, we compare the proposed BIM-based algorithm with two traditional optimization algorithms to investigate their impacts on repair effectiveness and computational overhead.

\end{compactitem}

\subsection{Studied Systems and Tasks}\label{sec:tasks}

Utilizing a public AI-CPS benchmark designed for industrial-level robotic manipulation~\cite{zhou2023towards}, we evaluate \abb{} across five different manipulation tasks. 
We use the Franka Emika robotic manipulator~\cite{Franka_Homepage} as the CPS and simulate the manipulation tasks with NVIDIA Omniverse Isaac Sim~\cite{isaac_requirements}, a physics engine-based simulation platform known for providing realistic simulations~\cite{makoviychuk2021isaac,zhou2023towards}.
Descriptions of the employed manipulation tasks are presented below (see also Fig.~\ref{fig:task_demo})

\begin{compactitem}[$\bullet$]
\item\textbf{Point Reaching (PR):} The robotic manipulator needs to reach a designated point with its end-effector. 
This basic skill is fundamental for completing more complex manipulation tasks.

\item\textbf{Cube Stacking (CS):} 
The robot must grasp a cube and place it on top of a target cube. 
This task requires the AI controller to accurately manage the spatial relationships between the two cubes.

\item\textbf{Peg in Hole (PH):} The robot is required to insert a peg into a small hole at the center of a pillar. 
This task demands precise control of the pose of the object.

\item\textbf{Ball Balancing (BB):} The robotic manipulator should hold a tray and keep a ball centred on the tray's top surface. 
Since the robot has no direct contact with the ball, this task introduces increased complexity in controller design.

\item\textbf{Ball Catching (BC):} The robot aims to catch a ball thrown at it using a box-shaped tool. 
This task requires the ability to interact with a moving object.
\end{compactitem}

For each task, we define the safety requirement as successfully completing the task's objective and describe this requirement using the STL specifications presented in~\cite{zhou2023towards}. 
These STL specifications are treated as standard specifications. 
Moreover, for the PR, CS, and BB tasks, we also implement a stricter STL specification by reducing the distance tolerance required for successful task completion. 
For example, in the CS task, the tolerance distance between the manipulated cube and the target cube is reduced from 0.024m to 0.012m. 
However, for the PH and BC tasks, where task completion is defined by whether the peg is successfully inserted or the ball is caught, reducing the distance tolerance does not yield meaningful results. 
Thus, stricter specifications are not applicable to these tasks.
Details of all employed standard and strict STL specifications are provided in Table~\ref{table:STL_specifiction}.

\begin{table}[t]
    \centering
    \caption{STL specifications for each considered task ($\square$: Globally; $\Diamond$: Eventually).}
    \label{table:STL_specifiction}
    \begin{tabular}{ccc}
    \toprule
         Task & Standard &  Strict \\
         \midrule
         PR & $\Diamond(\| \mathrm{finger}_{pos}-\mathrm{target}_{pos}\| \leq 0.24)$ & $\Diamond(\| \mathrm{finger}_{pos}-\mathrm{target}_{pos}\| \leq 0.06)$\\
         CS & \makecell[c]{$\Diamond (\|\mathrm{cube}_{pos}-\mathrm{target}_{pos}\| \leq 0.024$,\\ $\|\mathrm{cube}_z-\mathrm{target}_z\| \geq 0)$} & \makecell[c]{$\Diamond (\|\mathrm{cube}_{pos}-\mathrm{target}_{pos}\| \leq 0.012$,\\ $\|\mathrm{cube_z}-\mathrm{target}_z\| \geq 0)$}\\
         PH & $\square_{[4,5]}(\|\mathrm{peg}_{pos}-\mathrm{hole}_{pos}\|\leq0.12)$  &  not applicable\\
         BB & $\square_{[1,5]}(\|\mathrm{ball}_{pos}-\mathrm{tray}_{pos}\|\leq 0.25)$ & $\square_{[1,5]}(\|\mathrm{ball}_{pos}-\mathrm{tray}_{pos}\|\leq 0.1)$\\
         BC & $\Diamond(\|\mathrm{ball}_{pos}-\mathrm{tool}_{pos}\|\leq 0.1)$ & not applicable\\
         \bottomrule
    \end{tabular}
\end{table}

We employ two DRL control policies, trained using the PPO and TRPO algorithms, for each task. 
As noted in Section~\ref{sec:approach:data_collection}, well-trained policies typically result in primarily safe trajectories, indicating successful task completions. 
As shown in Table~\ref{table:policy_success}, these well-trained policies often achieve a success rate higher than 90\% under standard specifications. 
However, under strict specifications, the success rate drops, providing a greater opportunity to assess the effectiveness of \abb{} in enhancing the safety of AI controllers.

\begin{table}[t]
\centering
\caption{Success rate of the DRL control policies with respect to standard and strict specifications.}
 
\label{table:policy_success}
\begin{tabular}{lccccc}
\toprule
    Policy & PR & CS & PH & BB & BC\\
    \midrule
    PPO w.r.t standard & 0.99 & 1.00 & 0.82 & 0.98 & 1.00\\
    PPO w.r.t strict & 0.66 & 0.86 & - & 0.21 & -\\
    \midrule
    TRPO w.r.t standard & 1.00 & 0.99 & 0.92 & 1.00 & 0.97\\
    TRPO w.r.t strict & 0.71 & 0.80 & - &0.99 &-\\
\bottomrule
\end{tabular}
\end{table}

\subsection{Experimental Setup}

To minimize randomness in the experiments, for each task and DRL control policy, we use three different random seeds and report the average values. 
Detailed designs for each RQ are described below.

\subsubsection{RQ1}\label{sec:experiment_setting_rq1}

As mentioned in Section~\ref{sec:approach:data_collection}, we introduce action noises to DRL control policies during data collection to overcome the imbalance between safe and unsafe trajectories. 
The success rates of the noised policies, compared to the original policies, under standard specifications are presented in Table~\ref{table:collected_stats}. 
The results demonstrate an improved balance between safe and unsafe trajectories due to the injected noises.
In this work, we collect 5000 trajectories for each task and DRL controller, with each trajectory consisting of 300 timesteps. 
These datasets, collected with action noises and standard specifications, are then used to train the prediction models.

\begin{table}[t]
\centering
\caption{Success rate of the original and noised policies for data collection under standard specifications.}
\label{table:collected_stats}
 
\begin{tabular}{lccccc}
\toprule
    Policy & PR & CS & PH & BB & BC\\
    \midrule
    PPO & 0.99 & 1.00 & 0.82 & 0.98 & 1.00\\
    PPO with Noise & 0.59 & 0.57 & 0.43 & 0.79 & 0.71\\
    \midrule
    TRPO & 1.00 & 0.99 & 0.92 & 1.00 & 0.97\\
    TRPO with Noise & 0.63 &0.68 & 0.69&0.63 &0.54\\
\bottomrule
\end{tabular}
 
\end{table}

To evaluate the accuracy of the prediction model, we divide the collected dataset into training, validation, and testing subsets in a 7:1:2 ratio. 
Considering the detection of unsafe actions as a binary classification problem, we utilize four metrics to assess the effectiveness of the prediction model: accuracy, F1-score, mean square error (MSE), and the area under the receiver operating characteristic curves (AUC). 
The threshold $\psi_{thres}$ for the classification is set to 0 for all tasks.
While the accuracy and the F1-score measure the precision of the binary classification, MSE and AUC provide insights into the numerical accuracy of the prediction scores.

\subsubsection{RQ2}\label{sec:experiment_setting_rq2}
If an unsafe action is identified by the trained prediction model, we employ the targeted BIM to compute a repaired action. 
In our experiments, we set the step size and maximum iterations of the BIM to $\epsilon=0.1$ and $iter=3$, respectively. 
Note that, the noised policies are used only for collecting training data for the prediction model. 
To evaluate the effectiveness of \abb{}, we conduct experiments using the original DRL control policies.
We also compare \abb{} with Safe Exploration (SafeExp)~\cite{dalal2018safe}, a representative action repair method that uses an offline-trained safety layer to rectify erroneous actions. 
For more details about SafeExp, please refer to~\cite{dalal2018safe}.
To analyze the overhead introduced by \abb{}, we record the computational time when \abb{} is activated and compare it to the computational time when no action repair scheme is utilized.

\subsubsection{RQ3}
For examining the influence of optimization algorithms on the effectiveness of \abb{}, we compare the proposed BIM-based method with two traditional optimization algorithms: Nelder-Mead~\cite{gao2012implementing} and COBYLA~\cite{powell1994direct}. 
The performance of these algorithms heavily depends on the maximum iterations allowed. 
To ensure a fair comparison, we test these traditional algorithms at two different settings for maximum iterations. 
The first setting limits the iterations to 3, matching the setup for the targeted BIM. 
The second setting allows for 100 iterations, giving these algorithms more opportunity to find better solutions. 
We then assess and compare the repair performance and computational overhead under these different optimization algorithms.

\noindent\textbf{Software and Hardware Dependencies.}
All experiments are conducted with NVIDIA Omniverse Isaac Sim version 2022.2.0 on a machine equipped with an Intel i7-11800H CPU, NVIDIA 3080 GPU, and 64GB RAM.

\subsection{Evaluation Results}

In this subsection, we present detailed evaluation results for each RQ considered.

\subsubsection{\textbf{\rqone{}}\\}

\begin{table}[t]
\centering
\caption{RQ1 - The performance of the trained prediction models.}
\label{table:proxy_performance}

\begin{subtable}{\columnwidth}
\centering
\begin{tabular}{lcccc}
\toprule
Task & Accuracy & F1-score & MSE & AUC\\
\midrule
Point Reaching &  0.917   &    0.902  & 0.034 & 0.97\\
Cube Stacking &  0.910   &      0.909  & 0.087 & 0.97 \\
Peg In Hole  &  0.868   &     0.877   & 0.194 & 0.94 \\
Ball Balancing &  0.855   &      0.829  & 0.176  & 0.91\\
Ball Catching&  0.830   &   0.826 &  0.166 & 0.91\\
\bottomrule
\end{tabular}
\caption{PPO}
\end{subtable}

\begin{subtable}{\columnwidth}
\centering
\begin{tabular}{lcccc}
\toprule
Task & Accuracy & F1-score & MSE & AUC\\
\midrule
Point Reaching &  0.907   & 0.889     & 0.223 & 0.88\\
Cube Stacking &  0.870   &  0.804      & 0.187 & 0.89 \\
Peg In Hole  &  0.879   &   0.862    & 0.234 & 0.89 \\
Ball Balancing &   0.907  &  0.881   & 0.160 &0.93\\
Ball Catching&  0.941   &  0.941  & 0.025 &0.98 \\
\bottomrule
\end{tabular}
\caption{TRPO}
\end{subtable}
 
\end{table}

Table~\ref{table:proxy_performance} shows the performance of the trained prediction models for each manipulation task with PPO and TRPO control policies.

It can be observed that, for all prediction models, the classification accuracy ranges from 0.83 to 0.94, and the F1-score varies from 0.80 to 0.94.
Such high values indicate that the prediction model is capable of making accurate predictions, enabling an efficient determination of unsafe actions.

In all experiments, the prediction models also exhibit satisfying discriminatory capacity, with AUC scores exceeding $0.88$. 
Furthermore, MSE values are consistently low across all tasks, especially for the Point Reaching task with the PPO controller and the Ball Catching task with the TRPO controller, where MSE values are below $0.04$.
These metrics confirm the prediction models' satisfactory numerical accuracy and their effectiveness in guiding the action repair process.

\begin{ansbox}
    \textbf{Answer to RQ1:} 
    The trained prediction models are able to accurately predict the safety of actions and effectively guide the online action repair process.
\end{ansbox}

\begin{table*}[ht]
\centering
\caption{RQ2 - The performance of \abb{} for each manipulation task with PPO and TRPO control policies under standard and strict specifications. Values are the success rates. (best results are highlighted with \colorbox{\mygray}{gray})}
\label{table:rq2_res}
\begin{tabular}{c|ccc|ccc}
\toprule
   \multirow{2}{*}{Task} & \multicolumn{3}{c}{PPO w.r.t standard spec} & \multicolumn{3}{c}{PPO w.r.t strict spec}   \\
    & Ori        & SafeExp & \abb{} & Ori        & SafeExp & \abb{}\\ 
   \midrule
   Point Reaching & 0.99 &0.97 & \colorit{}1.00    & 0.66 &  0.60   & \colorit{}0.69   \\
   Cube Stacking & \colorit{}1.00 & \colorit{}1.00  & \colorit{}1.00   & 0.86    & 0.83       & \colorit{}0.92  \\
Peg In Hole   & 0.82       &  0.74          & \colorit{}0.86    & - & - &-   \\
Ball Balancing   & 0.98 & 0.98 &\colorit{}0.99   & 0.21 & 0.20    & \colorit{}0.29     \\
Ball Catching      & \colorit{}1.00          &  0.70          & \colorit{}1.00    & - & - &-         \\
\midrule
& \multicolumn{3}{c}{TRPO w.r.t standard spec}  & \multicolumn{3}{c}{TRPO w.r.t strict spec}\\
& Ori       & SafeExp & \abb & Ori       & SafeExp & \abb     \\
\midrule
Point Reaching & \colorit{}1.00 & \colorit{}1.00 & \colorit{}1.00 & 0.71   &  0.21       &  \colorit{}0.73\\
Cube Stacking& \colorit{}1.00 & 0.99 & \colorit{}1.00 & 0.80 & \colorit{}0.86   & 0.83 \\
Peg In Hole& 0.92      & 0.94          &  \colorit{}0.98  & - & - &-    \\
Ball Balancing &  \colorit{}1.00  & \colorit{}0.99 &   0.93 & \colorit{}0.99& \colorit{}1.00   & 0.99\\
Ball Catching  & 0.97      &  0.60           &  \colorit{}0.99 &- & - &-\\

\bottomrule
\end{tabular}
\end{table*}

\subsubsection{\textbf{\rqtwo{}}\\}
\label{sec:ans_rq2} 

\begin{table}[ht]
    \centering
    \caption{RQ2 - Average computational time (ms) of \abb{} per step.  Results are collected using the PPO control policy.}
    \label{table:rq2_overhead} 
     
    \begin{tabular}{c|cc|c}
    \toprule
         Task & Ori & \abb{} & Overhead\\
         \midrule
         Point Reaching  & 5.5  & 7.3 & 1.8 \\ 
         Cube Stacking & 8.1 & 9.5 & 1.4 \\ 
         Peg In Hole & 8.6 & 10.5 & 1.9 \\ 
         Ball Balancing & 6.5 & 7.7 & 1.2  \\ 
         Ball Catching & 8.7 & 9.2 & 0.5 \\
        \midrule 
        Average &  7.5 & 8.8 & 1.3 \\ 
        \bottomrule
    \end{tabular}
     
\end{table}

Table~\ref{table:rq2_res} showcases the performance of \abb{} in enhancing the safety of DRL control policies. 
Under standard specifications, \abb{} achieves the highest success rates across all tasks and control policies. 
Even for already well-performing DRL policies, \abb{} can still increase the success rates to nearly 100\%. 
When stricter specifications are applied, an expected decrease in overall success rates is observed. 
However, \abb{} is still able to enhance safety, achieving improved task completions in 5 out of 6 tasks. 
The only exception is the Cube Stacking task with the TRPO controller under the strict specification, where SafeExp performs best.
Nonetheless, the broad enhancement of safety across almost all tasks demonstrates the effectiveness of \abb{} in increasing the safety of DRL controllers, even when relying solely on input and output information for the action repair.

It is worth noting that SafeExp performs poorly in repairing actions for complex CPSs and tasks. 
One possible reason could be that SafeExp relies on creating a linear system model to estimate the consequences of each proposed action~\cite{dalal2018safe}. 
However, accurately modelling complex systems and tasks with a linear approach is often challenging and requires a considerable amount of training data. 
This limitation hinders the performance of SafeExp in the considered robotic manipulation tasks, which feature high-dimensional state spaces and highly nonlinear system dynamics.

Table~\ref{table:rq2_overhead} presents the analysis of the computational overhead for \abb{}. 
The results show that \abb{} requires an average of 8.8 milliseconds per action repair step, resulting in a computational time increase of approximately 1.3 milliseconds compared to the original control policy without action repair. 
This minimal increase still allows for real-time operation of the AI-CPSs, indicating that no latency issues are introduced and \abb{} effectively meets the real-time requirements.

\begin{ansbox}
    \textbf{Answer to RQ2:} 
    \abb{} is able to effectively repair unsafe actions across the examined AI-CPSs and tasks. 
    Meanwhile, it only adds minimal computational overhead, ensuring the real-time operation of the AI-CPSs.
\end{ansbox}

\begin{table*}[t]
    \centering
    \caption{RQ3 - The performance of different optimization algorithms in action repair. Values are the success rates (best results are highlighted with \colorbox{\mygray}{gray}). The Nelder-Mead (NM) and COBYLA (CO) algorithms are examined with 3 (NM(3), CO(3)) and 100 (NM(100), CO(100)) iterations, respectively. 
    }
    \label{table:rq3_repair_performance}
     
    \begin{tabular}{c|cccccc}
    \toprule
         \multirow{2}{*}{Task} &\multicolumn{6}{c}{PPO w.r.t standard spec}\\
         & Ori & \abb{} & NM(3) & NM(100) & CO(3) & CO(100)\\
         \midrule
         Point Reaching & 0.99 & \colorit{}1.00  & 0.99 & \colorit{}1.00 & 0.99 & 0.99 \\
         Cube Stacking & \colorit{}1.00 &\colorit{}1.00 & \colorit{}1.00 & \colorit{}1.00 & \colorit{}1.00 & \colorit{}1.00 \\
         Peg In Hole & 0.82 & \colorit{}0.86 & 0.74 & 0.68 & 0.78 & 0.75 \\
         Ball Balancing & 0.98 & \colorit{}0.99 & \colorit{}0.99  &\colorit{}0.99 & 0.97 & 0.96 \\
         Ball Catching & \colorit{}1.00& \colorit{}1.00& \colorit{}1.00& \colorit{}1.00& \colorit{}1.00& \colorit{}1.00\\
         \hline
         & \multicolumn{6}{c}{PPO w.r.t strict spec}\\
         & Ori & \abb{} & NM(3) & NM(100) & CO(3) & CO(100) \\
         \midrule
         Point Reaching & 0.66 & 0.69 & 0.68 & \colorit{}0.75 & 0.66 & 0.60 \\
         Cube Stacking & 0.86 & \colorit{}0.92 & 0.87 & 0.82 & 0.87 & 0.87\\
         Ball Balancing& 0.21 & \colorit{}0.29 & 0.24 & 0.24 & 0.27 & \colorit{}0.29\\
         \hline
         &\multicolumn{6}{c}{TRPO w.r.t standard spec} \\
         & Ori & \abb{} & NM(3) & NM(100) & CO(3) & CO(100)\\
         \midrule
         Point Reaching & \colorit{}1.00& \colorit{}1.00& \colorit{}1.00& \colorit{}1.00& \colorit{}1.00& \colorit{}1.00 \\
         Cube Stacking & \colorit{}1.00& \colorit{}1.00& \colorit{}1.00& 0.99& 0.99& \colorit{}1.00 \\
         Peg In Hole & 0.92 & \colorit{}0.98 & 0.91 & 0.93 & 0.92 & 0.95\\
         Ball Balancing & \colorit{}1.00& \colorit{}1.00& \colorit{}1.00& \colorit{}1.00& 0.99& \colorit{}1.00 \\
         Ball Catching & 0.97 & 0.99 & 0.95 & \colorit{}1.00 & 0.97 & 0.99 \\
         \hline
         & \multicolumn{6}{c}{TRPO w.r.t strict spec} \\
         & Ori & \abb{} & NM(3) & NM(100) & CO(3) & CO(100) \\
         \midrule
         Point Reaching & 0.71 & 0.73 & 0.69 & \colorit{}0.74 & 0.72 & 0.67 \\
         Cube Stacking & 0.80 & 0.83 & 0.85 & \colorit{}0.90 & 0.78 & 0.87\\
         Ball Balancing& \colorit{}0.99 & \colorit{}0.99 & 0.97 & \colorit{}0.99 & 0.97 &0.93\\
         \bottomrule
    \end{tabular}
\end{table*}

\begin{table}[!ht]
    \centering
    \caption{RQ3 - Average computational time (ms) per step for different optimization algorithms. Results are collected using the PPO control policy.}
    \label{table:rq3_overhead}
     
    \begin{tabular}{c|c|c|cc|cc}
    \toprule
         Task & Ori & \abb{} & NM(3) & NM(100) &CO(3) & CO(100)\\
         \midrule
         Point Reaching & 5.5  & 7.3 & 9.5 &  20.0 & 6.4 &  16.7 \\
         Cube Stacking & 8.1 & 9.5 & 9.2 &  30.7 & 9.2& 22.7\\
         Peg In Hole& 8.6 & 10.5 & 13.4 &  33.5 & 10.8 & 20.4\\
         Ball Balancing& 6.5 & 7.7 & 11.7 &  32.4 & 8.5&  21.5\\
        Ball Catching & 8.7 & 9.2 & 10.7 &  24.1 & 8.9 & 9.5\\
        \midrule
        Average & 7.3 & 8.8 & 10.9 & 28.1 & 8.8 & 18.2\\
        \bottomrule
    \end{tabular}
    \vspace{-15pt}
\end{table}

\subsubsection{\textbf{\rqthree{}}\\}\label{sec:ans_rq3} 

Table~\ref{table:rq3_repair_performance} displays the performance of different optimization algorithms in repairing unsafe actions. 
With only 3 iterations, the proposed targeted BIM-based algorithm outperforms the other two traditional optimization methods in 15 out of 16 tasks.
The only exception is the Cube Stacking task with the TRPO control policy under strict specifications, where Nelder-Mead achieves a slightly better result.
When the maximum iteration limit is increased to 100, the traditional optimization methods are able to find better solutions, leading to success rates comparable to our proposed method.

However, using a higher number of maximum iterations notably increases the computational overhead. 
As shown in Table~\ref{table:rq3_overhead}, running the Nelder-Mead and COBYLA algorithms with 100 maximum iterations typically results in computational times exceeding 20 ms. 
Given that a typical AI-CPS often operates the controller at 60 Hz (16.67 ms per step), such extended computational times can introduce delays in system operation and result in latency issues. 
Therefore, when considering both the effectiveness of action repair and computational efficiency, our proposed method provides the most satisfactory performance.

\begin{ansbox}
    \textbf{Answer to RQ3:} Compared to traditional optimization methods, our proposed targeted BIM-based algorithm achieves satisfactory performance in repairing unsafe actions while keeping computational overhead to a minimal level.
\end{ansbox}

\section{Discussion}\label{sec:discuss}

\noindent \textbf{Unsafe Action Detection and Repair using a Single Prediction Model.} 
In general, addressing the runtime safety issues of DRL control policies encompasses two primary tasks: detecting unsafe actions~\cite{gu2022review,zolfagharian2023smarla} and repairing these actions~\cite{zhou2020runtime,bharadhwaj2020conservative,bloem2015shield}. 
\abb{} connects these two tasks by leveraging the forecasting capability of the same prediction model. 
During runtime, we utilize the prediction model's forward inference to identify unsafe actions. 
Subsequently, in the repair process, we employ the same model's backpropagation to gather gradient information, guiding the generation of repaired actions. 
By integrating these two tasks through the dual utilization of the same prediction model, \abb{} ensures that the repair process is directly informed by the prediction model, thereby maintaining coherence between detection and repair. 

\noindent \textbf{Limitation.}
A major limitation of \abb{} is the prediction model's strong reliance on the specific combination of task and policy. 
The prediction model can only be applied to the same CPS task and policy for which it was trained, necessitating new training datasets for different policies, even within the same task. 
A potential solution to enhance generalizability across various policies could involve developing a prediction model that relies solely on the characteristics of the CPS. 
Such a policy-independent prediction model could then be seamlessly integrated into \abb{}, increasing its generalizability. 
Nevertheless, as an exploratory work, \abb{} has demonstrated satisfactory performance and promising potential in leveraging the same prediction model for both runtime monitoring and repairing.

\noindent \textbf{Future Direction.}

Based on the evaluation results, we intend to explore the following two directions in our future research:
\begin{compactitem}[$\bullet$]
    \item The performance of \abb{} heavily relies on the accuracy of the prediction model since it plays a key role in both unsafe action detection and following repair phases.
    Therefore, a crucial research direction is enhancing the accuracy and reliability of the prediction model, particularly in addressing false positive predictions, where unsafe actions are mistakenly predicted as safe. 
    \item In \abb{}, we perform a repaired action at each timestep where the unsafe action is detected. 
    However, for tasks requiring continuous time control, generating a series of repaired actions over a continuous time horizon may yield better results. 
    Developing such a repair scheme could potentially enhance the performance of \abb{}.
    \item In \abb{}, the state vector is an extracted one-dimensional vector. However, our framework has no design to handle tasks with multimedia data as input, such as robots equipped with camera sensors~\cite{bahl2020neural,singh2019end}. Therefore, if future developments allow for integrating perception modules~\cite{garg2020semantics} and state extraction algorithms~\cite{abel2018state} into \abb{}, its potential applications would be significantly broadened.
\end{compactitem}
\section{Threats to Validity}\label{sec:threats}

\noindent\textbf{Internal Validity.}
The quality of the prediction model can be an internal factor that affects the effectiveness of our runtime repair framework, as it not only triggers the repair process but also provides gradient information for generating repaired actions.
To mitigate this threat, we first add noises to the original policies to balance the distribution of safe and unsafe data.
Then, we leverage multiple metrics, i.e., accuracy, F1-score, MSE and AUC, to quantitatively assess the quality of the prediction model.

\noindent\textbf{External Validity.}
An external threat could be the generalizability of the proposed method to CPSs and DRL control policies other than those studied in this paper.
Considering that different CPSs and DRL policies may have various operation objectives, safety requirements, and behavior characteristics, it is crucial for \abb{} to be adaptable across diverse scenarios.
To mitigate this threat, we collect diverse CPSs with different operational goals and safety specifications.
Two representative DRL policies, i.e., PPO and TRPO, are examined with each CPS and task to demonstrate the effectiveness of our approach.
Given that \abb{} operates under a black-box runtime repair scheme, we consider our method capable of adapting to different CPS and DRL settings.
Another threat is the randomness inherent in the robot's movement and interaction with the environment. 
To mitigate the effects of this randomness, we simulate each experiment three times using different seeds and average them to obtain the final results.

\noindent\textbf{Construct Validity.}
One construct threat is that the metrics we applied in the evaluation may not comprehensively assess the effectiveness of the proposed repair framework.
To mitigate this threat, we evaluate our method from two aspects: the success rate of completing the tasks without safety violations and the computational overhead introduced by the repair strategy.
Namely, the former measures the effectiveness of \abb{} in correcting the faculty actions and enhancing the system safety, while the latter ensures that the repair process does not impose excessive overhead that could disrupt the real-time control process of the CPSs.

\section{Related Works}\label{sec:relatedWork}

\noindent\textbf{Deep Neural Network Repair.}
In terms of Deep Neural Network (DNN), the corresponding repair techniques can be generally classified into two categories as follows: 

\textit{Adversarial training} refers to methods that first collect adversarial examples (inputs deliberately perturbed to trigger DNN's faulty behaviors) and then use these examples to augment the training dataset and retrain the DNN model~\cite{balunovic2020adversarial, jia2022adversarial, shafahi2020universal, shafahi2019adversarial, bai2021recent}.
The key point of such approaches is to elevate the robustness of the DNN model against specific attacks by injecting the corresponding adversarial examples into the training loop.

\textit{Parameter alternation} implies the techniques that directly modify the parameters within the DNN to inherently correct the behavior of the model in response to adversarial inputs~\cite{zhang2019apricot, wang2019repairing, sohn2019search,tian2018deeptest,gong2022curiosity}. 
These methods generally require localizing the suspicious DNN parameters associated with faculty outputs and then optimizing these parameters to accomplish the desired repair goal.

Our method differs from these commonly used solutions in that we conduct the repairs in a black-box manner, which does not directly manipulate the subject model. 
Hence, our framework is considered to have better generalizability and adaptability.

\noindent\textbf{DRL Policy Repair.}
Unlike traditional DNN models, which mainly focus on classification and regression tasks, DRL policies are usually designed for decision-making tasks, such as planning and system control.
Therefore, repairing DRL policies is often more challenging, as the repair process needs to address not only the high-dimensional, continuous observation and action spaces but also the intricate system dynamics.
In the realm of reinforcement learning, one subfield which explicitly focuses on safeguarding the safety of DRL policies is Safe Reinforcement Learning (SRL)~\cite{alshiekh2018safe, garcia2015comprehensive, gu2022review, zanon2020safe, lutjens2019safe, zhou2020general, zhou2021learning}.

A common approach in SRL to improve safety involves modeling the system as a Constrained Markov Decision Process (CMDP)~\cite{altman2021constrained} and then learning a critic online to estimate the associated cost function~\cite{gu2022review,achiam2017constrained,yang2020projection}. 
This cost function then guides the policy update, steering it towards the direction that reduces the expected accumulated costs over a prediction horizon.
For example, Constrained Policy Optimization (CPO)~\cite{achiam2017constrained} used surrogate functions to approximate both objective and constraint functions in CMDP and updated the policy accordingly. 
Building on CPO, \cite{yang2020projection} introduced Projection-based Constrained Policy Optimization (PCPO), a two-step method that first solves the policy search via TRPO and then projects the policy back to a feasible region to satisfy safety constraints.
Another direction of SRL is to first adjust the data collection method based on the characteristics of the actual task, and then update the policy through retraining.
For instance, Lyu et al.~\cite{lyu2023autorepair} proposed a framework that involves collecting trajectory data, using a signal diagnosis tool to identify potential risk periods. 
Then, it generated repaired trajectories using a genetic algorithm to create safety-assured datasets for retraining the policy.
Compared to these approaches, \abb{} does not estimate a cost function or require policy retraining. 
Instead, it focuses on online rectification of potentially dangerous actions without modifying the original DRL policy.

To address the high effort and computational costs associated with policy retraining, several action repair methods have also been proposed~\cite{alshiekh2018safe,fisac2018general,zhou2020general,zhou2021learning, dalal2018safe}.
For instance, Alshiekh et al.~\cite{alshiekh2018safe} proposed a reactive system, \textit{shield}, to safeguard the actions of an RL policy at both the learning and execution phases. 
It utilized an abstract model to actively monitor the system's safety and correct any necessary actions.
However, such an approach proves challenging for CPSs, as the high-dimensional state and action spaces complicate the development of a scalable abstract model to identify unsafe states.
Alternatively, in \abb{}, instead of constructing an abstract model, we utilize a DNN model to predict the STL scores of the CPS, which can adequately reflect the safety property of the system at runtime. 
Moreover, Dalal et al.~\cite{dalal2018safe} introduced Safe Exploration, which created a linear model of the system and then concatenated a safety layer on top of the original policy.
We take inspiration from this work, but instead of deriving an analytic solution to obtain safe actions, \abb{} generates the repaired actions through a BIM-based optimization that is guided by the predicted STL scores.
\section{Conclusion}
\label{sec:conclusion}

In this work, we introduce \abb{}, a runtime repair framework designed for AI-CPSs with DRL controllers. 
For a given CPS task and an associated DRL control policy, \abb{} first constructs a prediction model to forecast the safety of each action proposed by the policy.
Such a model serves two primary functions: it acts as a runtime monitor to identify unsafe actions and initiate the repair process, and it provides safety estimates and gradient information to guide the action repair. 
The generation of repaired actions is achieved by using a targeted BIM-based optimization strategy, which is able to efficiently find rectified actions that drive the system towards a safe condition. 
Meanwhile, \abb{} also keeps the computational overhead at a minimal level, ensuring the real-time operation of the AI-CPSs. 
Extensive experiments are conducted with various CPS tasks, DRL policies and optimization algorithms to validate the effectiveness of \abb{}.
The results show that \abb{} can effectively repair faculty outputs from the original DRL policy and improve the overall safety and reliability of the AI-CPSs.

\bibliographystyle{ACM-Reference-Format}
\bibliography{ref}


\begin{thebibliography}{73}


\ifx \showCODEN    \undefined \def \showCODEN     #1{\unskip}     \fi
\ifx \showDOI      \undefined \def \showDOI       #1{#1}\fi
\ifx \showISBNx    \undefined \def \showISBNx     #1{\unskip}     \fi
\ifx \showISBNxiii \undefined \def \showISBNxiii  #1{\unskip}     \fi
\ifx \showISSN     \undefined \def \showISSN      #1{\unskip}     \fi
\ifx \showLCCN     \undefined \def \showLCCN      #1{\unskip}     \fi
\ifx \shownote     \undefined \def \shownote      #1{#1}          \fi
\ifx \showarticletitle \undefined \def \showarticletitle #1{#1}   \fi
\ifx \showURL      \undefined \def \showURL       {\relax}        \fi
\providecommand\bibfield[2]{#2}
\providecommand\bibinfo[2]{#2}
\providecommand\natexlab[1]{#1}
\providecommand\showeprint[2][]{arXiv:#2}

\bibitem[Fra(2024)]%
        {Franka_Homepage}
 \bibinfo{year}{2024}\natexlab{}.
\newblock \bibinfo{title}{Franka Emika Website}.
\newblock \bibinfo{howpublished}{\url{https://franka.de/}}.
\newblock


\bibitem[isa(2024)]%
        {isaac_requirements}
 \bibinfo{year}{2024}\natexlab{}.
\newblock \bibinfo{title}{Isaac Sim}.
\newblock \bibinfo{howpublished}{\url{ttps://docs.omniverse.nvidia.com/isaacsim}}.
\newblock


\bibitem[Abel et~al\mbox{.}(2018)]%
        {abel2018state}
\bibfield{author}{\bibinfo{person}{David Abel}, \bibinfo{person}{Dilip Arumugam}, \bibinfo{person}{Lucas Lehnert}, {and} \bibinfo{person}{Michael Littman}.} \bibinfo{year}{2018}\natexlab{}.
\newblock \showarticletitle{State abstractions for lifelong reinforcement learning}. In \bibinfo{booktitle}{\emph{International Conference on Machine Learning}}. PMLR, \bibinfo{pages}{10--19}.
\newblock


\bibitem[Achiam et~al\mbox{.}(2017)]%
        {achiam2017constrained}
\bibfield{author}{\bibinfo{person}{Joshua Achiam}, \bibinfo{person}{David Held}, \bibinfo{person}{Aviv Tamar}, {and} \bibinfo{person}{Pieter Abbeel}.} \bibinfo{year}{2017}\natexlab{}.
\newblock \showarticletitle{Constrained policy optimization}. In \bibinfo{booktitle}{\emph{International Conference on Machine Learning}}. PMLR, \bibinfo{pages}{22--31}.
\newblock


\bibitem[Afzal et~al\mbox{.}(2020)]%
        {afzal2020study}
\bibfield{author}{\bibinfo{person}{Afsoon Afzal}, \bibinfo{person}{Deborah~S Katz}, \bibinfo{person}{Claire~Le Goues}, {and} \bibinfo{person}{Christopher~S Timperley}.} \bibinfo{year}{2020}\natexlab{}.
\newblock \showarticletitle{A study on the challenges of using robotics simulators for testing}.
\newblock \bibinfo{journal}{\emph{arXiv preprint arXiv:2004.07368}} (\bibinfo{year}{2020}).
\newblock


\bibitem[Alguliyev et~al\mbox{.}(2018)]%
        {alguliyev2018cyber}
\bibfield{author}{\bibinfo{person}{Rasim Alguliyev}, \bibinfo{person}{Yadigar Imamverdiyev}, {and} \bibinfo{person}{Lyudmila Sukhostat}.} \bibinfo{year}{2018}\natexlab{}.
\newblock \showarticletitle{Cyber-physical systems and their security issues}.
\newblock \bibinfo{journal}{\emph{Computers in Industry}}  \bibinfo{volume}{100} (\bibinfo{year}{2018}), \bibinfo{pages}{212--223}.
\newblock


\bibitem[Alshiekh et~al\mbox{.}(2018)]%
        {alshiekh2018safe}
\bibfield{author}{\bibinfo{person}{Mohammed Alshiekh}, \bibinfo{person}{Roderick Bloem}, \bibinfo{person}{R{\"u}diger Ehlers}, \bibinfo{person}{Bettina K{\"o}nighofer}, \bibinfo{person}{Scott Niekum}, {and} \bibinfo{person}{Ufuk Topcu}.} \bibinfo{year}{2018}\natexlab{}.
\newblock \showarticletitle{Safe reinforcement learning via shielding}. In \bibinfo{booktitle}{\emph{Proceedings of the AAAI conference on artificial intelligence}}, Vol.~\bibinfo{volume}{32}.
\newblock


\bibitem[Altman(2021)]%
        {altman2021constrained}
\bibfield{author}{\bibinfo{person}{Eitan Altman}.} \bibinfo{year}{2021}\natexlab{}.
\newblock \bibinfo{booktitle}{\emph{Constrained Markov decision processes}}.
\newblock \bibinfo{publisher}{Routledge}.
\newblock


\bibitem[Ang et~al\mbox{.}(2005)]%
        {ang2005pid}
\bibfield{author}{\bibinfo{person}{Kiam~Heong Ang}, \bibinfo{person}{Gregory Chong}, {and} \bibinfo{person}{Yun Li}.} \bibinfo{year}{2005}\natexlab{}.
\newblock \showarticletitle{PID control system analysis, design, and technology}.
\newblock \bibinfo{journal}{\emph{IEEE transactions on control systems technology}} \bibinfo{volume}{13}, \bibinfo{number}{4} (\bibinfo{year}{2005}), \bibinfo{pages}{559--576}.
\newblock


\bibitem[Bahl et~al\mbox{.}(2020)]%
        {bahl2020neural}
\bibfield{author}{\bibinfo{person}{Shikhar Bahl}, \bibinfo{person}{Mustafa Mukadam}, \bibinfo{person}{Abhinav Gupta}, {and} \bibinfo{person}{Deepak Pathak}.} \bibinfo{year}{2020}\natexlab{}.
\newblock \showarticletitle{Neural dynamic policies for end-to-end sensorimotor learning}.
\newblock \bibinfo{journal}{\emph{Advances in Neural Information Processing Systems}}  \bibinfo{volume}{33} (\bibinfo{year}{2020}), \bibinfo{pages}{5058--5069}.
\newblock


\bibitem[Bai et~al\mbox{.}(2021)]%
        {bai2021recent}
\bibfield{author}{\bibinfo{person}{Tao Bai}, \bibinfo{person}{Jinqi Luo}, \bibinfo{person}{Jun Zhao}, \bibinfo{person}{Bihan Wen}, {and} \bibinfo{person}{Qian Wang}.} \bibinfo{year}{2021}\natexlab{}.
\newblock \showarticletitle{Recent advances in adversarial training for adversarial robustness}.
\newblock \bibinfo{journal}{\emph{arXiv preprint arXiv:2102.01356}} (\bibinfo{year}{2021}).
\newblock


\bibitem[Balunovi{\'c} and Vechev(2020)]%
        {balunovic2020adversarial}
\bibfield{author}{\bibinfo{person}{Mislav Balunovi{\'c}} {and} \bibinfo{person}{Martin Vechev}.} \bibinfo{year}{2020}\natexlab{}.
\newblock \showarticletitle{Adversarial training and provable defenses: Bridging the gap}. In \bibinfo{booktitle}{\emph{8th International Conference on Learning Representations (ICLR 2020)(virtual)}}. International Conference on Learning Representations.
\newblock


\bibitem[Bartocci et~al\mbox{.}(2018)]%
        {bartocci2018specification}
\bibfield{author}{\bibinfo{person}{Ezio Bartocci}, \bibinfo{person}{Jyotirmoy Deshmukh}, \bibinfo{person}{Alexandre Donz{\'e}}, \bibinfo{person}{Georgios Fainekos}, \bibinfo{person}{Oded Maler}, \bibinfo{person}{Dejan Ni{\v{c}}kovi{\'c}}, {and} \bibinfo{person}{Sriram Sankaranarayanan}.} \bibinfo{year}{2018}\natexlab{}.
\newblock \showarticletitle{Specification-based monitoring of cyber-physical systems: a survey on theory, tools and applications}.
\newblock \bibinfo{journal}{\emph{Lectures on Runtime Verification: Introductory and Advanced Topics}} (\bibinfo{year}{2018}), \bibinfo{pages}{135--175}.
\newblock


\bibitem[Bazaraa et~al\mbox{.}(2013)]%
        {bazaraa2013nonlinear}
\bibfield{author}{\bibinfo{person}{Mokhtar~S Bazaraa}, \bibinfo{person}{Hanif~D Sherali}, {and} \bibinfo{person}{Chitharanjan~M Shetty}.} \bibinfo{year}{2013}\natexlab{}.
\newblock \bibinfo{booktitle}{\emph{Nonlinear programming: theory and algorithms}}.
\newblock \bibinfo{publisher}{John wiley \& sons}.
\newblock


\bibitem[Bharadhwaj et~al\mbox{.}(2020)]%
        {bharadhwaj2020conservative}
\bibfield{author}{\bibinfo{person}{Homanga Bharadhwaj}, \bibinfo{person}{Aviral Kumar}, \bibinfo{person}{Nicholas Rhinehart}, \bibinfo{person}{Sergey Levine}, \bibinfo{person}{Florian Shkurti}, {and} \bibinfo{person}{Animesh Garg}.} \bibinfo{year}{2020}\natexlab{}.
\newblock \showarticletitle{Conservative Safety Critics for Exploration}. In \bibinfo{booktitle}{\emph{International Conference on Learning Representations}}.
\newblock


\bibitem[Bloem et~al\mbox{.}(2015)]%
        {bloem2015shield}
\bibfield{author}{\bibinfo{person}{Roderick Bloem}, \bibinfo{person}{Bettina K{\"o}nighofer}, \bibinfo{person}{Robert K{\"o}nighofer}, {and} \bibinfo{person}{Chao Wang}.} \bibinfo{year}{2015}\natexlab{}.
\newblock \showarticletitle{Shield synthesis: Runtime enforcement for reactive systems}. In \bibinfo{booktitle}{\emph{International conference on tools and algorithms for the construction and analysis of systems}}. Springer, \bibinfo{pages}{533--548}.
\newblock


\bibitem[Castiglioni et~al\mbox{.}(2024)]%
        {castiglioni2024stark}
\bibfield{author}{\bibinfo{person}{Valentina Castiglioni}, \bibinfo{person}{Michele Loreti}, {and} \bibinfo{person}{Simone Tini}.} \bibinfo{year}{2024}\natexlab{}.
\newblock \showarticletitle{Stark: a tool for the analysis of CPSs robustness}.
\newblock \bibinfo{journal}{\emph{Science of Computer Programming}} (\bibinfo{year}{2024}), \bibinfo{pages}{103134}.
\newblock


\bibitem[Dalal et~al\mbox{.}(2018)]%
        {dalal2018safe}
\bibfield{author}{\bibinfo{person}{Gal Dalal}, \bibinfo{person}{Krishnamurthy Dvijotham}, \bibinfo{person}{Matej Vecerik}, \bibinfo{person}{Todd Hester}, \bibinfo{person}{Cosmin Paduraru}, {and} \bibinfo{person}{Yuval Tassa}.} \bibinfo{year}{2018}\natexlab{}.
\newblock \showarticletitle{Safe exploration in continuous action spaces}.
\newblock \bibinfo{journal}{\emph{arXiv preprint arXiv:1801.08757}} (\bibinfo{year}{2018}).
\newblock


\bibitem[Donz{\'e}(2010)]%
        {donze2010breach}
\bibfield{author}{\bibinfo{person}{Alexandre Donz{\'e}}.} \bibinfo{year}{2010}\natexlab{}.
\newblock \showarticletitle{Breach, a toolbox for verification and parameter synthesis of hybrid systems.}. In \bibinfo{booktitle}{\emph{CAV}}, Vol.~\bibinfo{volume}{10}. Springer, \bibinfo{pages}{167--170}.
\newblock


\bibitem[Donz{\'e} and Maler(2010)]%
        {donze2010robust}
\bibfield{author}{\bibinfo{person}{Alexandre Donz{\'e}} {and} \bibinfo{person}{Oded Maler}.} \bibinfo{year}{2010}\natexlab{}.
\newblock \showarticletitle{Robust satisfaction of temporal logic over real-valued signals}. In \bibinfo{booktitle}{\emph{International Conference on Formal Modeling and Analysis of Timed Systems}}. Springer, \bibinfo{pages}{92--106}.
\newblock


\bibitem[Fisac et~al\mbox{.}(2019)]%
        {fisac2018general}
\bibfield{author}{\bibinfo{person}{Jaime~F Fisac}, \bibinfo{person}{Anayo~K Akametalu}, \bibinfo{person}{Melanie~N Zeilinger}, \bibinfo{person}{Shahab Kaynama}, \bibinfo{person}{Jeremy Gillula}, {and} \bibinfo{person}{Claire~J Tomlin}.} \bibinfo{year}{2019}\natexlab{}.
\newblock \showarticletitle{A general safety framework for learning-based control in uncertain robotic systems}.
\newblock \bibinfo{journal}{\emph{IEEE Trans. Automat. Control}} \bibinfo{volume}{64}, \bibinfo{number}{7} (\bibinfo{date}{Jul.} \bibinfo{year}{2019}), \bibinfo{pages}{2737--2752}.
\newblock


\bibitem[Gao and Han(2012)]%
        {gao2012implementing}
\bibfield{author}{\bibinfo{person}{Fuchang Gao} {and} \bibinfo{person}{Lixing Han}.} \bibinfo{year}{2012}\natexlab{}.
\newblock \showarticletitle{Implementing the Nelder-Mead simplex algorithm with adaptive parameters}.
\newblock \bibinfo{journal}{\emph{Computational Optimization and Applications}} \bibinfo{volume}{51}, \bibinfo{number}{1} (\bibinfo{year}{2012}), \bibinfo{pages}{259--277}.
\newblock


\bibitem[Garc{\i}a and Fern{\'a}ndez(2015)]%
        {garcia2015comprehensive}
\bibfield{author}{\bibinfo{person}{Javier Garc{\i}a} {and} \bibinfo{person}{Fernando Fern{\'a}ndez}.} \bibinfo{year}{2015}\natexlab{}.
\newblock \showarticletitle{A comprehensive survey on safe reinforcement learning}.
\newblock \bibinfo{journal}{\emph{Journal of Machine Learning Research}} \bibinfo{volume}{16}, \bibinfo{number}{1} (\bibinfo{year}{2015}), \bibinfo{pages}{1437--1480}.
\newblock


\bibitem[Garg et~al\mbox{.}(2020)]%
        {garg2020semantics}
\bibfield{author}{\bibinfo{person}{Sourav Garg}, \bibinfo{person}{Niko S{\"u}nderhauf}, \bibinfo{person}{Feras Dayoub}, \bibinfo{person}{Douglas Morrison}, \bibinfo{person}{Akansel Cosgun}, \bibinfo{person}{Gustavo Carneiro}, \bibinfo{person}{Qi Wu}, \bibinfo{person}{Tat-Jun Chin}, \bibinfo{person}{Ian Reid}, \bibinfo{person}{Stephen Gould}, {et~al\mbox{.}}} \bibinfo{year}{2020}\natexlab{}.
\newblock \showarticletitle{Semantics for robotic mapping, perception and interaction: A survey}.
\newblock \bibinfo{journal}{\emph{Foundations and Trends{\textregistered} in Robotics}} \bibinfo{volume}{8}, \bibinfo{number}{1--2} (\bibinfo{year}{2020}), \bibinfo{pages}{1--224}.
\newblock


\bibitem[Gong et~al\mbox{.}(2022)]%
        {gong2022curiosity}
\bibfield{author}{\bibinfo{person}{Chen Gong}, \bibinfo{person}{Zhou Yang}, \bibinfo{person}{Yunpeng Bai}, \bibinfo{person}{Jieke Shi}, \bibinfo{person}{Arunesh Sinha}, \bibinfo{person}{Bowen Xu}, \bibinfo{person}{David Lo}, \bibinfo{person}{Xinwen Hou}, {and} \bibinfo{person}{Guoliang Fan}.} \bibinfo{year}{2022}\natexlab{}.
\newblock \showarticletitle{Curiosity-driven and victim-aware adversarial policies}. In \bibinfo{booktitle}{\emph{Proceedings of the 38th Annual Computer Security Applications Conference}}. \bibinfo{pages}{186--200}.
\newblock


\bibitem[Goodfellow et~al\mbox{.}(2014)]%
        {goodfellow2014explaining}
\bibfield{author}{\bibinfo{person}{Ian~J Goodfellow}, \bibinfo{person}{Jonathon Shlens}, {and} \bibinfo{person}{Christian Szegedy}.} \bibinfo{year}{2014}\natexlab{}.
\newblock \showarticletitle{Explaining and harnessing adversarial examples}.
\newblock \bibinfo{journal}{\emph{arXiv preprint arXiv:1412.6572}} (\bibinfo{year}{2014}).
\newblock


\bibitem[Gu et~al\mbox{.}(2022)]%
        {gu2022review}
\bibfield{author}{\bibinfo{person}{Shangding Gu}, \bibinfo{person}{Long Yang}, \bibinfo{person}{Yali Du}, \bibinfo{person}{Guang Chen}, \bibinfo{person}{Florian Walter}, \bibinfo{person}{Jun Wang}, \bibinfo{person}{Yaodong Yang}, {and} \bibinfo{person}{Alois Knoll}.} \bibinfo{year}{2022}\natexlab{}.
\newblock \showarticletitle{A review of safe reinforcement learning: Methods, theory and applications}.
\newblock \bibinfo{journal}{\emph{arXiv preprint arXiv:2205.10330}} (\bibinfo{year}{2022}).
\newblock


\bibitem[Hu et~al\mbox{.}(2023)]%
        {hu2023AIoTML}
\bibfield{author}{\bibinfo{person}{Ming Hu}, \bibinfo{person}{E. Cao}, \bibinfo{person}{Hongbing Huang}, \bibinfo{person}{Min Zhang}, \bibinfo{person}{Xiaohong Chen}, {and} \bibinfo{person}{Mingsong Chen}.} \bibinfo{year}{2023}\natexlab{}.
\newblock \showarticletitle{AIoTML: A Unified Modeling Language for AIoT-Based Cyber–Physical Systems}.
\newblock \bibinfo{journal}{\emph{IEEE Transactions on Computer-Aided Design of Integrated Circuits and Systems}} \bibinfo{volume}{42}, \bibinfo{number}{11} (\bibinfo{year}{2023}), \bibinfo{pages}{3545--3558}.
\newblock
\urldef\tempurl%
\url{https://doi.org/10.1109/TCAD.2023.3264786}
\showDOI{\tempurl}


\bibitem[Hu et~al\mbox{.}(2020)]%
        {Hu2020Quantitative}
\bibfield{author}{\bibinfo{person}{Ming Hu}, \bibinfo{person}{Wenxue Duan}, \bibinfo{person}{Min Zhang}, \bibinfo{person}{Tongquan Wei}, {and} \bibinfo{person}{Mingsong Chen}.} \bibinfo{year}{2020}\natexlab{}.
\newblock \showarticletitle{Quantitative Timing Analysis for Cyber-Physical Systems Using Uncertainty-Aware Scenario-Based Specifications}.
\newblock \bibinfo{journal}{\emph{IEEE Transactions on Computer-Aided Design of Integrated Circuits and Systems}} \bibinfo{volume}{39}, \bibinfo{number}{11} (\bibinfo{year}{2020}), \bibinfo{pages}{4006--4017}.
\newblock
\urldef\tempurl%
\url{https://doi.org/10.1109/TCAD.2020.3012843}
\showDOI{\tempurl}


\bibitem[Jia et~al\mbox{.}(2021)]%
        {jia2021integrated}
\bibfield{author}{\bibinfo{person}{Dongyao Jia}, \bibinfo{person}{Jie Sun}, \bibinfo{person}{Anshuman Sharma}, \bibinfo{person}{Zuduo Zheng}, {and} \bibinfo{person}{Bingyi Liu}.} \bibinfo{year}{2021}\natexlab{}.
\newblock \showarticletitle{Integrated simulation platform for conventional, connected and automated driving: A design from cyber--physical systems perspective}.
\newblock \bibinfo{journal}{\emph{Transportation Research Part C: Emerging Technologies}}  \bibinfo{volume}{124} (\bibinfo{year}{2021}), \bibinfo{pages}{102984}.
\newblock


\bibitem[Jia et~al\mbox{.}(2022)]%
        {jia2022adversarial}
\bibfield{author}{\bibinfo{person}{Xiaojun Jia}, \bibinfo{person}{Yong Zhang}, \bibinfo{person}{Baoyuan Wu}, \bibinfo{person}{Ke Ma}, \bibinfo{person}{Jue Wang}, {and} \bibinfo{person}{Xiaochun Cao}.} \bibinfo{year}{2022}\natexlab{}.
\newblock \showarticletitle{LAS-AT: adversarial training with learnable attack strategy}. In \bibinfo{booktitle}{\emph{Proceedings of the IEEE/CVF Conference on Computer Vision and Pattern Recognition}}. \bibinfo{pages}{13398--13408}.
\newblock


\bibitem[Kurakin et~al\mbox{.}(2018)]%
        {kurakin2018adversarial}
\bibfield{author}{\bibinfo{person}{Alexey Kurakin}, \bibinfo{person}{Ian~J Goodfellow}, {and} \bibinfo{person}{Samy Bengio}.} \bibinfo{year}{2018}\natexlab{}.
\newblock \showarticletitle{Adversarial examples in the physical world}.
\newblock In \bibinfo{booktitle}{\emph{Artificial intelligence safety and security}}. \bibinfo{publisher}{Chapman and Hall/CRC}, \bibinfo{pages}{99--112}.
\newblock


\bibitem[Lee(2015)]%
        {lee2015past}
\bibfield{author}{\bibinfo{person}{Edward~A Lee}.} \bibinfo{year}{2015}\natexlab{}.
\newblock \showarticletitle{The past, present and future of cyber-physical systems: A focus on models}.
\newblock \bibinfo{journal}{\emph{Sensors}} \bibinfo{volume}{15}, \bibinfo{number}{3} (\bibinfo{year}{2015}), \bibinfo{pages}{4837--4869}.
\newblock


\bibitem[Leong et~al\mbox{.}(2020)]%
        {leong2020deep}
\bibfield{author}{\bibinfo{person}{Alex~S Leong}, \bibinfo{person}{Arunselvan Ramaswamy}, \bibinfo{person}{Daniel~E Quevedo}, \bibinfo{person}{Holger Karl}, {and} \bibinfo{person}{Ling Shi}.} \bibinfo{year}{2020}\natexlab{}.
\newblock \showarticletitle{Deep reinforcement learning for wireless sensor scheduling in cyber--physical systems}.
\newblock \bibinfo{journal}{\emph{Automatica}}  \bibinfo{volume}{113} (\bibinfo{year}{2020}), \bibinfo{pages}{108759}.
\newblock


\bibitem[Li et~al\mbox{.}(2020)]%
        {li2020early}
\bibfield{author}{\bibinfo{person}{Nianyu Li}, \bibinfo{person}{Christos Tsigkanos}, \bibinfo{person}{Zhi Jin}, \bibinfo{person}{Zhenjiang Hu}, {and} \bibinfo{person}{Carlo Ghezzi}.} \bibinfo{year}{2020}\natexlab{}.
\newblock \showarticletitle{Early validation of cyber--physical space systems via multi-concerns integration}.
\newblock \bibinfo{journal}{\emph{Journal of Systems and Software}}  \bibinfo{volume}{170} (\bibinfo{year}{2020}), \bibinfo{pages}{110742}.
\newblock


\bibitem[Lillicrap et~al\mbox{.}(2015)]%
        {lillicrap2015continuous}
\bibfield{author}{\bibinfo{person}{Timothy~P Lillicrap}, \bibinfo{person}{Jonathan~J Hunt}, \bibinfo{person}{Alexander Pritzel}, \bibinfo{person}{Nicolas Heess}, \bibinfo{person}{Tom Erez}, \bibinfo{person}{Yuval Tassa}, \bibinfo{person}{David Silver}, {and} \bibinfo{person}{Daan Wierstra}.} \bibinfo{year}{2015}\natexlab{}.
\newblock \showarticletitle{Continuous control with deep reinforcement learning}.
\newblock \bibinfo{journal}{\emph{arXiv preprint arXiv:1509.02971}} (\bibinfo{year}{2015}).
\newblock


\bibitem[Lin et~al\mbox{.}(2020)]%
        {lin2020comparison}
\bibfield{author}{\bibinfo{person}{Yuan Lin}, \bibinfo{person}{John McPhee}, {and} \bibinfo{person}{Nasser~L Azad}.} \bibinfo{year}{2020}\natexlab{}.
\newblock \showarticletitle{Comparison of deep reinforcement learning and model predictive control for adaptive cruise control}.
\newblock \bibinfo{journal}{\emph{IEEE Transactions on Intelligent Vehicles}} \bibinfo{volume}{6}, \bibinfo{number}{2} (\bibinfo{year}{2020}), \bibinfo{pages}{221--231}.
\newblock


\bibitem[Liu et~al\mbox{.}(2019)]%
        {liu2019reinforcement}
\bibfield{author}{\bibinfo{person}{Xing Liu}, \bibinfo{person}{Hansong Xu}, \bibinfo{person}{Weixian Liao}, {and} \bibinfo{person}{Wei Yu}.} \bibinfo{year}{2019}\natexlab{}.
\newblock \showarticletitle{Reinforcement learning for cyber-physical systems}. In \bibinfo{booktitle}{\emph{2019 IEEE International Conference on Industrial Internet (ICII)}}. IEEE, \bibinfo{pages}{318--327}.
\newblock


\bibitem[L{\"u}tjens et~al\mbox{.}(2019)]%
        {lutjens2019safe}
\bibfield{author}{\bibinfo{person}{Bj{\"o}rn L{\"u}tjens}, \bibinfo{person}{Michael Everett}, {and} \bibinfo{person}{Jonathan~P How}.} \bibinfo{year}{2019}\natexlab{}.
\newblock \showarticletitle{Safe reinforcement learning with model uncertainty estimates}. In \bibinfo{booktitle}{\emph{2019 International Conference on Robotics and Automation (ICRA)}}. IEEE, \bibinfo{pages}{8662--8668}.
\newblock


\bibitem[Lv et~al\mbox{.}(2021)]%
        {lv2021artificial}
\bibfield{author}{\bibinfo{person}{Zhihan Lv}, \bibinfo{person}{Dongliang Chen}, \bibinfo{person}{Ranran Lou}, {and} \bibinfo{person}{Ammar Alazab}.} \bibinfo{year}{2021}\natexlab{}.
\newblock \showarticletitle{Artificial intelligence for securing industrial-based cyber--physical systems}.
\newblock \bibinfo{journal}{\emph{Future generation computer systems}}  \bibinfo{volume}{117} (\bibinfo{year}{2021}), \bibinfo{pages}{291--298}.
\newblock


\bibitem[Lyu et~al\mbox{.}(2023)]%
        {lyu2023autorepair}
\bibfield{author}{\bibinfo{person}{Deyun Lyu}, \bibinfo{person}{Jiayang Song}, \bibinfo{person}{Zhenya Zhang}, \bibinfo{person}{Zhijie Wang}, \bibinfo{person}{Tianyi Zhang}, \bibinfo{person}{Lei Ma}, {and} \bibinfo{person}{Jianjun Zhao}.} \bibinfo{year}{2023}\natexlab{}.
\newblock \showarticletitle{AutoRepair: Automated Repair for AI-Enabled Cyber-Physical Systems under Safety-Critical Conditions}.
\newblock \bibinfo{journal}{\emph{arXiv preprint arXiv:2304.05617}} (\bibinfo{year}{2023}).
\newblock


\bibitem[Makoviychuk et~al\mbox{.}(2021)]%
        {makoviychuk2021isaac}
\bibfield{author}{\bibinfo{person}{Viktor Makoviychuk}, \bibinfo{person}{Lukasz Wawrzyniak}, \bibinfo{person}{Yunrong Guo}, \bibinfo{person}{Michelle Lu}, \bibinfo{person}{Kier Storey}, \bibinfo{person}{Miles Macklin}, \bibinfo{person}{David Hoeller}, \bibinfo{person}{Nikita Rudin}, \bibinfo{person}{Arthur Allshire}, \bibinfo{person}{Ankur Handa}, {et~al\mbox{.}}} \bibinfo{year}{2021}\natexlab{}.
\newblock \showarticletitle{Isaac gym: High performance gpu-based physics simulation for robot learning}.
\newblock \bibinfo{journal}{\emph{arXiv preprint arXiv:2108.10470}} (\bibinfo{year}{2021}).
\newblock


\bibitem[Morari et~al\mbox{.}(1988)]%
        {morari1988model}
\bibfield{author}{\bibinfo{person}{Manfred Morari}, \bibinfo{person}{Carlos~E Garcia}, {and} \bibinfo{person}{David~M Prett}.} \bibinfo{year}{1988}\natexlab{}.
\newblock \showarticletitle{Model predictive control: theory and practice}.
\newblock \bibinfo{journal}{\emph{IFAC Proceedings Volumes}} \bibinfo{volume}{21}, \bibinfo{number}{4} (\bibinfo{year}{1988}), \bibinfo{pages}{1--12}.
\newblock


\bibitem[Nikolakis et~al\mbox{.}(2019)]%
        {nikolakis2019cyber}
\bibfield{author}{\bibinfo{person}{Nikolaos Nikolakis}, \bibinfo{person}{Vasilis Maratos}, {and} \bibinfo{person}{Sotiris Makris}.} \bibinfo{year}{2019}\natexlab{}.
\newblock \showarticletitle{A cyber physical system (CPS) approach for safe human-robot collaboration in a shared workplace}.
\newblock \bibinfo{journal}{\emph{Robotics and Computer-Integrated Manufacturing}}  \bibinfo{volume}{56} (\bibinfo{year}{2019}), \bibinfo{pages}{233--243}.
\newblock


\bibitem[Plaat et~al\mbox{.}(2020)]%
        {plaat2020deep}
\bibfield{author}{\bibinfo{person}{Aske Plaat}, \bibinfo{person}{Walter Kosters}, {and} \bibinfo{person}{Mike Preuss}.} \bibinfo{year}{2020}\natexlab{}.
\newblock \showarticletitle{Deep model-based reinforcement learning for high-dimensional problems, a survey}.
\newblock \bibinfo{journal}{\emph{arXiv preprint arXiv:2008.05598}} (\bibinfo{year}{2020}).
\newblock


\bibitem[Powell(1994)]%
        {powell1994direct}
\bibfield{author}{\bibinfo{person}{Michael~JD Powell}.} \bibinfo{year}{1994}\natexlab{}.
\newblock \bibinfo{booktitle}{\emph{A direct search optimization method that models the objective and constraint functions by linear interpolation}}.
\newblock \bibinfo{publisher}{Springer}.
\newblock


\bibitem[Radanliev et~al\mbox{.}(2021)]%
        {radanliev2021artificial}
\bibfield{author}{\bibinfo{person}{Petar Radanliev}, \bibinfo{person}{David De~Roure}, \bibinfo{person}{Max Van~Kleek}, \bibinfo{person}{Omar Santos}, {and} \bibinfo{person}{Uchenna Ani}.} \bibinfo{year}{2021}\natexlab{}.
\newblock \showarticletitle{Artificial intelligence in cyber physical systems}.
\newblock \bibinfo{journal}{\emph{AI \& society}}  \bibinfo{volume}{36} (\bibinfo{year}{2021}), \bibinfo{pages}{783--796}.
\newblock


\bibitem[Rungger and Tabuada(2015)]%
        {rungger2015notion}
\bibfield{author}{\bibinfo{person}{Matthias Rungger} {and} \bibinfo{person}{Paulo Tabuada}.} \bibinfo{year}{2015}\natexlab{}.
\newblock \showarticletitle{A notion of robustness for cyber-physical systems}.
\newblock \bibinfo{journal}{\emph{IEEE Trans. Automat. Control}} \bibinfo{volume}{61}, \bibinfo{number}{8} (\bibinfo{year}{2015}), \bibinfo{pages}{2108--2123}.
\newblock


\bibitem[Schulman et~al\mbox{.}(2015)]%
        {schulman2015trust}
\bibfield{author}{\bibinfo{person}{John Schulman}, \bibinfo{person}{Sergey Levine}, \bibinfo{person}{Pieter Abbeel}, \bibinfo{person}{Michael Jordan}, {and} \bibinfo{person}{Philipp Moritz}.} \bibinfo{year}{2015}\natexlab{}.
\newblock \showarticletitle{Trust region policy optimization}. In \bibinfo{booktitle}{\emph{International conference on machine learning}}. PMLR, \bibinfo{pages}{1889--1897}.
\newblock


\bibitem[Schulman et~al\mbox{.}(2017)]%
        {schulman2017proximal}
\bibfield{author}{\bibinfo{person}{John Schulman}, \bibinfo{person}{Filip Wolski}, \bibinfo{person}{Prafulla Dhariwal}, \bibinfo{person}{Alec Radford}, {and} \bibinfo{person}{Oleg Klimov}.} \bibinfo{year}{2017}\natexlab{}.
\newblock \showarticletitle{Proximal policy optimization algorithms}.
\newblock \bibinfo{journal}{\emph{arXiv preprint arXiv:1707.06347}} (\bibinfo{year}{2017}).
\newblock


\bibitem[Shafahi et~al\mbox{.}(2019)]%
        {shafahi2019adversarial}
\bibfield{author}{\bibinfo{person}{Ali Shafahi}, \bibinfo{person}{Mahyar Najibi}, \bibinfo{person}{Mohammad~Amin Ghiasi}, \bibinfo{person}{Zheng Xu}, \bibinfo{person}{John Dickerson}, \bibinfo{person}{Christoph Studer}, \bibinfo{person}{Larry~S Davis}, \bibinfo{person}{Gavin Taylor}, {and} \bibinfo{person}{Tom Goldstein}.} \bibinfo{year}{2019}\natexlab{}.
\newblock \showarticletitle{Adversarial training for free!}
\newblock \bibinfo{journal}{\emph{Advances in neural information processing systems}}  \bibinfo{volume}{32} (\bibinfo{year}{2019}).
\newblock


\bibitem[Shafahi et~al\mbox{.}(2020)]%
        {shafahi2020universal}
\bibfield{author}{\bibinfo{person}{Ali Shafahi}, \bibinfo{person}{Mahyar Najibi}, \bibinfo{person}{Zheng Xu}, \bibinfo{person}{John Dickerson}, \bibinfo{person}{Larry~S Davis}, {and} \bibinfo{person}{Tom Goldstein}.} \bibinfo{year}{2020}\natexlab{}.
\newblock \showarticletitle{Universal adversarial training}. In \bibinfo{booktitle}{\emph{Proceedings of the AAAI Conference on Artificial Intelligence}}, Vol.~\bibinfo{volume}{34}. \bibinfo{pages}{5636--5643}.
\newblock


\bibitem[Singh et~al\mbox{.}(2019)]%
        {singh2019end}
\bibfield{author}{\bibinfo{person}{Avi Singh}, \bibinfo{person}{Larry Yang}, \bibinfo{person}{Kristian Hartikainen}, \bibinfo{person}{Chelsea Finn}, {and} \bibinfo{person}{Sergey Levine}.} \bibinfo{year}{2019}\natexlab{}.
\newblock \showarticletitle{End-to-end robotic reinforcement learning without reward engineering}.
\newblock \bibinfo{journal}{\emph{arXiv preprint arXiv:1904.07854}} (\bibinfo{year}{2019}).
\newblock


\bibitem[Sohn et~al\mbox{.}(2019)]%
        {sohn2019search}
\bibfield{author}{\bibinfo{person}{Jeongju Sohn}, \bibinfo{person}{Sungmin Kang}, {and} \bibinfo{person}{Shin Yoo}.} \bibinfo{year}{2019}\natexlab{}.
\newblock \showarticletitle{Search based repair of deep neural networks}.
\newblock \bibinfo{journal}{\emph{arXiv preprint arXiv:1912.12463}} (\bibinfo{year}{2019}).
\newblock


\bibitem[Song et~al\mbox{.}(2016)]%
        {song2016cyber}
\bibfield{author}{\bibinfo{person}{Houbing~H Song}, \bibinfo{person}{Danda~B Rawat}, \bibinfo{person}{Sabina Jeschke}, {and} \bibinfo{person}{Christian Brecher}.} \bibinfo{year}{2016}\natexlab{}.
\newblock \bibinfo{booktitle}{\emph{Cyber-physical systems: foundations, principles and applications}}.
\newblock \bibinfo{publisher}{Morgan Kaufmann}.
\newblock


\bibitem[Song et~al\mbox{.}(2022)]%
        {song2022cyber}
\bibfield{author}{\bibinfo{person}{Jiayang Song}, \bibinfo{person}{Deyun Lyu}, \bibinfo{person}{Zhenya Zhang}, \bibinfo{person}{Zhijie Wang}, \bibinfo{person}{Tianyi Zhang}, {and} \bibinfo{person}{Lei Ma}.} \bibinfo{year}{2022}\natexlab{}.
\newblock \showarticletitle{When cyber-physical systems meet AI: a benchmark, an evaluation, and a way forward}. In \bibinfo{booktitle}{\emph{Proceedings of the 44th International Conference on Software Engineering: Software Engineering in Practice}}. \bibinfo{pages}{343--352}.
\newblock


\bibitem[Song et~al\mbox{.}(2023)]%
        {song2023mathtt}
\bibfield{author}{\bibinfo{person}{Jiayang Song}, \bibinfo{person}{Xuan Xie}, {and} \bibinfo{person}{Lei Ma}.} \bibinfo{year}{2023}\natexlab{}.
\newblock \showarticletitle{SIEGE: A Semantics-Guided Safety Enhancement Framework for AI-enabled Cyber-Physical Systems}.
\newblock \bibinfo{journal}{\emph{IEEE Transactions on Software Engineering}} (\bibinfo{year}{2023}).
\newblock


\bibitem[Tabuada et~al\mbox{.}(2014)]%
        {tabuada2014towards}
\bibfield{author}{\bibinfo{person}{Paulo Tabuada}, \bibinfo{person}{Sina~Yamac Caliskan}, \bibinfo{person}{Matthias Rungger}, {and} \bibinfo{person}{Rupak Majumdar}.} \bibinfo{year}{2014}\natexlab{}.
\newblock \showarticletitle{Towards robustness for cyber-physical systems}.
\newblock \bibinfo{journal}{\emph{IEEE Trans. Automat. Control}} \bibinfo{volume}{59}, \bibinfo{number}{12} (\bibinfo{year}{2014}), \bibinfo{pages}{3151--3163}.
\newblock


\bibitem[Tian et~al\mbox{.}(2018)]%
        {tian2018deeptest}
\bibfield{author}{\bibinfo{person}{Yuchi Tian}, \bibinfo{person}{Kexin Pei}, \bibinfo{person}{Suman Jana}, {and} \bibinfo{person}{Baishakhi Ray}.} \bibinfo{year}{2018}\natexlab{}.
\newblock \showarticletitle{Deeptest: Automated testing of deep-neural-network-driven autonomous cars}. In \bibinfo{booktitle}{\emph{Proceedings of the 40th international conference on software engineering}}. \bibinfo{pages}{303--314}.
\newblock


\bibitem[Wang et~al\mbox{.}(2019)]%
        {wang2019repairing}
\bibfield{author}{\bibinfo{person}{Hao Wang}, \bibinfo{person}{Berk Ustun}, {and} \bibinfo{person}{Flavio Calmon}.} \bibinfo{year}{2019}\natexlab{}.
\newblock \showarticletitle{Repairing without retraining: Avoiding disparate impact with counterfactual distributions}. In \bibinfo{booktitle}{\emph{International Conference on Machine Learning}}. PMLR, \bibinfo{pages}{6618--6627}.
\newblock


\bibitem[Xiao et~al\mbox{.}(2018)]%
        {xiao2018generating}
\bibfield{author}{\bibinfo{person}{Chaowei Xiao}, \bibinfo{person}{Bo Li}, \bibinfo{person}{Jun-Yan Zhu}, \bibinfo{person}{Warren He}, \bibinfo{person}{Mingyan Liu}, {and} \bibinfo{person}{Dawn Song}.} \bibinfo{year}{2018}\natexlab{}.
\newblock \showarticletitle{Generating adversarial examples with adversarial networks}.
\newblock \bibinfo{journal}{\emph{arXiv preprint arXiv:1801.02610}} (\bibinfo{year}{2018}).
\newblock


\bibitem[Xie et~al\mbox{.}(2023)]%
        {xie2023mosaic}
\bibfield{author}{\bibinfo{person}{Xuan Xie}, \bibinfo{person}{Jiayang Song}, \bibinfo{person}{Zhehua Zhou}, \bibinfo{person}{Fuyuan Zhang}, {and} \bibinfo{person}{Lei Ma}.} \bibinfo{year}{2023}\natexlab{}.
\newblock \showarticletitle{Mosaic: Model-based Safety Analysis Framework for AI-enabled Cyber-Physical Systems}.
\newblock \bibinfo{journal}{\emph{arXiv preprint arXiv:2305.03882}} (\bibinfo{year}{2023}).
\newblock


\bibitem[Yang et~al\mbox{.}(2020)]%
        {yang2020projection}
\bibfield{author}{\bibinfo{person}{Tsung-Yen Yang}, \bibinfo{person}{Justinian Rosca}, \bibinfo{person}{Karthik Narasimhan}, {and} \bibinfo{person}{Peter~J Ramadge}.} \bibinfo{year}{2020}\natexlab{}.
\newblock \showarticletitle{Projection-based constrained policy optimization}.
\newblock \bibinfo{journal}{\emph{ArXiv preprint arXiv:2010.03152}} (\bibinfo{year}{2020}).
\newblock


\bibitem[Yang et~al\mbox{.}(2022)]%
        {yang2022neural}
\bibfield{author}{\bibinfo{person}{Xiaodong Yang}, \bibinfo{person}{Tom Yamaguchi}, \bibinfo{person}{Hoang-Dung Tran}, \bibinfo{person}{Bardh Hoxha}, \bibinfo{person}{Taylor~T Johnson}, {and} \bibinfo{person}{Danil Prokhorov}.} \bibinfo{year}{2022}\natexlab{}.
\newblock \showarticletitle{Neural network repair with reachability analysis}. In \bibinfo{booktitle}{\emph{International Conference on Formal Modeling and Analysis of Timed Systems}}. Springer, \bibinfo{pages}{221--236}.
\newblock


\bibitem[Yohanandhan et~al\mbox{.}(2020)]%
        {yohanandhan2020cyber}
\bibfield{author}{\bibinfo{person}{Rajaa~Vikhram Yohanandhan}, \bibinfo{person}{Rajvikram~Madurai Elavarasan}, \bibinfo{person}{Premkumar Manoharan}, {and} \bibinfo{person}{Lucian Mihet-Popa}.} \bibinfo{year}{2020}\natexlab{}.
\newblock \showarticletitle{Cyber-physical power system (CPPS): A review on modeling, simulation, and analysis with cyber security applications}.
\newblock \bibinfo{journal}{\emph{IEEE Access}}  \bibinfo{volume}{8} (\bibinfo{year}{2020}), \bibinfo{pages}{151019--151064}.
\newblock


\bibitem[Yuan et~al\mbox{.}(2019)]%
        {yuan2019adversarial}
\bibfield{author}{\bibinfo{person}{Xiaoyong Yuan}, \bibinfo{person}{Pan He}, \bibinfo{person}{Qile Zhu}, {and} \bibinfo{person}{Xiaolin Li}.} \bibinfo{year}{2019}\natexlab{}.
\newblock \showarticletitle{Adversarial examples: Attacks and defenses for deep learning}.
\newblock \bibinfo{journal}{\emph{IEEE transactions on neural networks and learning systems}} \bibinfo{volume}{30}, \bibinfo{number}{9} (\bibinfo{year}{2019}), \bibinfo{pages}{2805--2824}.
\newblock


\bibitem[Zanon and Gros(2020)]%
        {zanon2020safe}
\bibfield{author}{\bibinfo{person}{Mario Zanon} {and} \bibinfo{person}{S{\'e}bastien Gros}.} \bibinfo{year}{2020}\natexlab{}.
\newblock \showarticletitle{Safe reinforcement learning using robust MPC}.
\newblock \bibinfo{journal}{\emph{IEEE Trans. Automat. Control}} \bibinfo{volume}{66}, \bibinfo{number}{8} (\bibinfo{year}{2020}), \bibinfo{pages}{3638--3652}.
\newblock


\bibitem[Zhang and Chan(2019)]%
        {zhang2019apricot}
\bibfield{author}{\bibinfo{person}{Hao Zhang} {and} \bibinfo{person}{WK Chan}.} \bibinfo{year}{2019}\natexlab{}.
\newblock \showarticletitle{Apricot: A weight-adaptation approach to fixing deep learning models}. In \bibinfo{booktitle}{\emph{2019 34th IEEE/ACM International Conference on Automated Software Engineering (ASE)}}. IEEE, \bibinfo{pages}{376--387}.
\newblock


\bibitem[Zhou et~al\mbox{.}(2020a)]%
        {zhou2020runtime}
\bibfield{author}{\bibinfo{person}{Weichao Zhou}, \bibinfo{person}{Ruihan Gao}, \bibinfo{person}{BaekGyu Kim}, \bibinfo{person}{Eunsuk Kang}, {and} \bibinfo{person}{Wenchao Li}.} \bibinfo{year}{2020}\natexlab{a}.
\newblock \showarticletitle{Runtime-safety-guided policy repair}. In \bibinfo{booktitle}{\emph{Runtime Verification: 20th International Conference, RV 2020, Los Angeles, CA, USA, October 6--9, 2020, Proceedings 20}}. Springer, \bibinfo{pages}{131--150}.
\newblock


\bibitem[Zhou et~al\mbox{.}(2020b)]%
        {zhou2020general}
\bibfield{author}{\bibinfo{person}{Zhehua Zhou}, \bibinfo{person}{Ozgur~S Oguz}, \bibinfo{person}{Marion Leibold}, {and} \bibinfo{person}{Martin Buss}.} \bibinfo{year}{2020}\natexlab{b}.
\newblock \showarticletitle{A general framework to increase safety of learning algorithms for dynamical systems based on region of attraction estimation}.
\newblock \bibinfo{journal}{\emph{IEEE Transactions on Robotics}} \bibinfo{volume}{36}, \bibinfo{number}{5} (\bibinfo{year}{2020}), \bibinfo{pages}{1472--1490}.
\newblock


\bibitem[Zhou et~al\mbox{.}(2021)]%
        {zhou2021learning}
\bibfield{author}{\bibinfo{person}{Zhehua Zhou}, \bibinfo{person}{Ozgur~S Oguz}, \bibinfo{person}{Marion Leibold}, {and} \bibinfo{person}{Martin Buss}.} \bibinfo{year}{2021}\natexlab{}.
\newblock \showarticletitle{Learning a low-dimensional representation of a safe region for safe reinforcement learning on dynamical systems}.
\newblock \bibinfo{journal}{\emph{IEEE Transactions on Neural Networks and Learning Systems}} \bibinfo{volume}{34}, \bibinfo{number}{5} (\bibinfo{year}{2021}), \bibinfo{pages}{2513--2527}.
\newblock


\bibitem[Zhou et~al\mbox{.}(2024)]%
        {zhou2023towards}
\bibfield{author}{\bibinfo{person}{Zhehua Zhou}, \bibinfo{person}{Jiayang Song}, \bibinfo{person}{Xuan Xie}, \bibinfo{person}{Zhan Shu}, \bibinfo{person}{Lei Ma}, \bibinfo{person}{Dikai Liu}, \bibinfo{person}{Jianxiong Yin}, {and} \bibinfo{person}{Simon See}.} \bibinfo{year}{2024}\natexlab{}.
\newblock \showarticletitle{Towards building ai-cps with nvidia isaac sim: An industrial benchmark and case study for robotics manipulation}. In \bibinfo{booktitle}{\emph{Proceedings of the 46th International Conference on Software Engineering: Software Engineering in Practice}}. \bibinfo{pages}{263--274}.
\newblock


\bibitem[Zolfagharian et~al\mbox{.}(2023)]%
        {zolfagharian2023smarla}
\bibfield{author}{\bibinfo{person}{Amirhossein Zolfagharian}, \bibinfo{person}{Manel Abdellatif}, \bibinfo{person}{Lionel~C Briand}, {et~al\mbox{.}}} \bibinfo{year}{2023}\natexlab{}.
\newblock \showarticletitle{SMARLA: A Safety Monitoring Approach for Deep Reinforcement Learning Agents}.
\newblock \bibinfo{journal}{\emph{arXiv preprint arXiv:2308.02594}} (\bibinfo{year}{2023}).
\newblock


\end{thebibliography}

\end{document}